\documentclass[final,11pt]{elsarticle}

\usepackage{amsmath}
\usepackage{amssymb}
\usepackage{amsfonts}
\usepackage{graphicx}
\usepackage{graphics}
\usepackage{textcomp}

\usepackage{scalerel}
\usepackage{tikz}
\usepackage{tikz-3dplot}
\usepackage[utf8]{inputenc}
\usepackage{pgfplots}
\pgfplotsset{compat=newest}
\usepgfplotslibrary{groupplots}
\usepgfplotslibrary{dateplot}
\usetikzlibrary{svg.path}
\definecolor{orcidlogocol}{HTML}{A6CE39}
\tikzset{
orcidlogo/.pic={
\fill[orcidlogocol] svg{M256,128c0,70.7-57.3,128-128,128C57.3,256,0,198.7,0,128C0,57.3,57.3,0,128,0C198.7,0,256,57.3,256,128z};
\fill[white] svg{M86.3,186.2H70.9V79.1h15.4v48.4V186.2z}
svg{M108.9,79.1h41.6c39.6,0,57,28.3,57,53.6c0,27.5-21.5,53.6-56.8,53.6h-41.8V79.1z M124.3,172.4h24.5c34.9,0,42.9-26.5,42.9-39.7c0-21.5-13.7-39.7-43.7-39.7h-23.7V172.4z}
svg{M88.7,56.8c0,5.5-4.5,10.1-10.1,10.1c-5.6,0-10.1-4.6-10.1-10.1c0-5.6,4.5-10.1,10.1-10.1C84.2,46.7,88.7,51.3,88.7,56.8z};
}
}
\newcommand\orcidicon[1]{\href{https://orcid.org/#1}{\mbox{\scalerel*{
\begin{tikzpicture}[yscale=-1,transform shape]
\pic{orcidlogo};
\end{tikzpicture}
}{|}}}}

\usepackage{hyperref}

\usepackage{url}
\usepackage{lineno,hyperref}
\usepackage{booktabs}
\usepackage[labelfont=bf]{caption}
\usepackage{threeparttable}
\usepackage{lscape}
\usepackage{afterpage}
\usepackage[dvips]{geometry}
\usepackage{setspace} 
\usepackage{multirow}

\usepackage{tikz}
\usetikzlibrary{shapes,arrows.meta,calc,fit,backgrounds,shapes.multipart,positioning,chains,decorations.shapes,mindmap,trees}

\definecolor{my_yellow}{RGB}{230, 229, 145}
\definecolor{my_blue}{RGB}{177, 223, 239}   
\definecolor{my_green}{RGB}{177, 235, 160}    
\definecolor{my_red}{RGB}{232, 151, 158}
\definecolor{my_orange}{RGB}{255, 192, 116}
\definecolor{my_gray}{RGB}{179, 182, 183}
\definecolor{my_purple}{RGB}{180,150,200}

\usepackage{amssymb}
\usepackage{pifont}
\newcommand{\xmark}{\ding{55}}
\definecolor{g1}{RGB}{146, 43, 33}

\makeatletter
\def\NAT@def@citea{\def\@citea{\NAT@separator}}
\makeatother

\biboptions{sort&compress}

\makeatletter
\renewcommand\paragraph{\@startsection{paragraph}{4}{\z@}%
            {-2.5ex\@plus -1ex \@minus -.25ex}%
            {1.25ex \@plus .25ex}%
            {\normalfont\normalsize\itshape}}
\makeatother
\usepackage{indentfirst}

\usepackage{ragged2e}

\usepackage{comment}

\usepackage{tabularx}
\newcolumntype{Y}{>{\centering\arraybackslash}X}
\newcolumntype{C}[1]{>{\centering\let\newline\\\arraybackslash\hspace{0pt}}m{#1}}
\usepackage{float}

\newcolumntype{H}[1]{>{\hsize=#1\hsize\arraybackslash}X}

\usepackage{smartdiagram}
\begin{document}

\begin{frontmatter}

\title{{\fontsize{13.5}{10}\selectfont\bfseries Ancillary\,Services\,in\,Power\,System\,Transition\,Toward\,a\,100\%\,Non-Fossil Future: Market Design Challenges in the United States and Europe}}

\author[1]{Luigi Viola}
\ead{luigi@dsee.fee.unicamp.br, https://orcid.org/0000-0001-7913-5685}
\author[2]{Saeed Nordin}
\ead{saeedmoh@kth.se, https://orcid.org/0000-0003-1823-9653}
\author[1]{Daniel Dotta}
\ead{dottad@unicamp.br, https://orcid.org/0000-0002-3287-172X}
\author[2]{Mohammad Reza Hesamzadeh\corref{cor1}}
\ead{mrhesa@kth.se, https://orcid.org/0000-0002-9998-9773}
\author[3]{Ross Baldick}
\ead{baldick@ece.utexas.edu, https://orcid.org/0000-0003-2783-7321}
\author[4]{Damian Flynn}
\ead{damian.flynn@ucd.ie, https://orcid.org/0000-0003-4638-9333}
{
\address[1]{University of Campinas, Av. Albert Einstein, 400, Campinas-SP, Brazil}
\address[2]{KTH Royal Institute of Technology, 100 44 Stockholm, Sweden}
\address[3]{University of Texas at Austin, Austin, TX 78705, USA}
\address[4]{University College Dublin, 4 Dublin, Ireland}
}

\cortext[cor1]{Corresponding author.}

\begin{abstract}

The expansion of variable generation has driven a transition toward a 100\% non-fossil power system. New system needs are challenging system stability and suggesting the need for a redesign of the ancillary service (AS) markets. This paper presents a comprehensive and broad review for industrial practitioners and academic researchers regarding the challenges and potential solutions to accommodate high shares of variable renewable energy (VRE) generation levels. We detail the main drivers enabling the energy transition and facilitating the provision of ASs. A systematic review of the United States and European AS markets is conducted. We clearly organize the main ASs in a standard taxonomy, identifying current practices and initiatives to support the increasing VRE share. Furthermore, we envision the future of modern AS markets, proposing potential solutions for some remaining fundamental technical and market design challenges.

\end{abstract}


\end{frontmatter}

\section*{Acknowledgements}
The National Agency of Electric Energy (ANEEL) research and development program has supported this work, with financial support provided by Engie (under grant PD-00403-0053/2021) and the National Council for Scientific and Technological Development (CNPq).

\section*{Abbreviations}

aFRR, Automatic Frequency Restoration Reserve;
AGC, Automatic Generation Control; 
AI, Artificial Intelligence;
AS, Ancillary Service; 
BESS, Battery Energy Storage Systems; 
BRP, Balance Responsible Party; 
BSC, Black Start Capability; 
BSP, Balance Service Provider; 
DAM, Day-Ahead-Market; 
DER, Distributed Energy Resource;
DG, Distributed Generation; 
DRR, Dynamic Reactive Response; 
DSO, Distribution System Operator; 
ESS, Energy Storage Systems; 
EU, Europe;
EV, Electric Vehicles;
FACTS, Flexible AC Transmission System; 
FCR, Frequency Containment Reserve; 
FFR, Fast Frequency Response; 
FPFAPR, Fast Post-Fault Active Power Recovery;
FR, Frequency Regulation; 
FRC, Flexible Ramping Capability; 
GFL, Grid-Following; 
GFM, Grid-Forming; 
HVDC, High Voltage Direct Current; 
IBR, Inverter-Based Resource;
ICT, Information and Communication Technology;
IDM, Intraday Market; 
IR, Inertial Response; 
ISO, Independent System Operator; 
LMP, Locational Marginal Prices; 
LOC, Lost Opportunity Cost;
mFRR, Manual Frequency Restoration Reserve; 
MP, Marginal Pricing; 
NSR, Non-Spinning Reserve; 
ORDC, Operating Reserve Demand Curve; 
PBP, Pay-as-Bid; 
PFC, Primary Frequency Control; 
PFR, Primary Frequency Response; 
PHS, Pumped Hydropower Storage;
PLL, Phase-Lock Loop; 
PMU, Phasor Measurement Unit;
PRF, Primary Frequency Response;
PV, Photovoltaic; 
RoCoF, Rate of Change of Frequency; 
RP, Regulated Price 
RR, Replacement Reserves; 
RTM, Real-Time Market; 
RUC, Reliability Unit Commitment;  
SC, Synchronous Condensers; 
SCED, Security-Constrained Economic Dispatch 
SCUC, Security-Constrained Unit Commitment;
SFC, Secondary Frequency Control;
SG,Synchronous Generator;
SIR, Synchronous Inertial Response;
SO, System Operator; 
SR, Spinning Reserve; 
SSRR, Steady-State Reactive Response;
STATCOM, Static Synchronous Compensators; 
SVC, Static VAR Compensators;
TFC, Tertiary Frequency Control; 
TSO, Transmission System Operator;
UFLS, Under-Frequency Load-Shedding; 
UPS, Uninterruptible Power Supply;
US, United States;
VIR, Virtual Inertial Response; 
VPP, Virtual Power Plant; 
VRE, Variable Renewable Energy;

\section{Introduction}

Power systems have witnessed a growing share of variable renewable energy (VRE) in the generation mix. This process is motivated by climate change concerns, aiming to reduce carbon emissions to limit global average temperature rises. By 2050, the United States (US) will reach 44\% of renewable electricity supply~\cite{EIA2022}. Currently, the regions operated by CAISO (California) and ERCOT (Texas) present significant solar and wind shares, respectively. The target in Europe (EU) is to reach 32\% renewable electricity supply in 2030~\cite{EU2020}. Great Britain (GB), Ireland (all-island), Germany, and the Nordic power system have dominant renewable generation participation. The path toward a 100\% non-fossil future includes a mix of VRE (wind and solar), other renewables (hydropower, ocean, and tidal), low-carbon sources such as biomass, geothermal, and nuclear power plants, and energy storage (pumped hydropower plants, batteries, hydrogen systems). Even if the average generation share of VRE in a power system is less than 100\%, power system stability problems may occur in a 100\% non-fossil future due to high VRE instantaneous share~\cite{Lew2020}.  Thus, traditional assumptions for power system operation based on synchronous generators (SGs) capabilities must be revisited, particularly the sufficient frequency and voltage support, and fuel availability (fossil-fuels-based units). A new set of system needs arises to preserve system stability and provide flexible operation when fewer synchronous resources are available. Inertia and fast response reserves are critical to maintain frequency stability. Voltage stability is impacted due to the scarcity of steady-state reactive power capability in regions with weaker transmission networks. The increased electrical distance between the remaining synchronous units requires enhanced dynamic reactive support. Additionally, ramping capability from flexible resources is essential to manage the variability and uncertainty of VRE~\cite{Nolan2019}.  

Historically, power systems have comprised large and centralized dispatchable SGs, mainly fossil-fueled (from coal, oil, and natural gas) or nuclear-, and hydro-powered units. System inertia and strength (measured by the short circuit ratio) are sufficiently supplied as a byproduct of SG operation. Thermal generation is capacity-constrained, has significant variable costs, and baseload units (coal-fired and nuclear plants) are inflexible. System operators (SOs) generally consider a unidirectional power flow from transmission through the distribution system, which is modeled as an aggregated load to reduce computational complexity. Distribution system operators (DSOs) may have limited ability to operate the network, relying on the capacity defined at the planning stage (i.e. fit-and-forget approach). 

In contrast to historical practice, the transition toward a 100\% non-fossil power system is boosted mostly by VRE generation and energy storage systems (ESS) interfaced by electronic inverters and located throughout the transmission and distribution system. Although centrally dispatched at the transmission level, the dispersion of non-dispatchable inverter-based resources (IBRs) through the distribution system tends to decentralize the generation. VRE is constrained by the availability of primary energy sources (i.e. energy-constrained) with very low variable costs (i.e. near zero-marginal cost). High shares of wind and solar generation result in a variable and uncertain net load profile~\cite{Mohandes2019} with steeper ramps and deeper valleys, requiring the dispatch of flexible resources. VRE displaces traditional generation in the merit-order dispatch and, combined with the inability of IBRs to provide an inertial response, reduces system inertia levels, shortening the response time to disturbances. Thus, fast-acting reserves must be dispatched to arrest the frequency drop following a disturbance, such as a generator tripping offline~\cite{Fernandez-Guillamon2019}. Also, tripping of online synchronous resources reduces reactive power capability and imposes new requirements to preserve voltage stability in response to small imbalances or contingencies. Increasing shares of distributed generation (DG) and other energy systems, such as batteries, requires enhanced load modeling to improve the visibility of distributed energy resources (DERs) in power system operation. Insufficient modeling of DERs in power flow studies may result in inefficient management of reactive power and congestion, also compromising contingency plans. Active operation of the distribution systems and close coordination between SOs and DSOs may unlock flexible resources, such as electric vehicles (EVs), heat pumps, dispersed generation and storage, improving system security.

Ancillary services (ASs) are crucial to help SOs in frequency response, voltage control, and system restoration to robustly ensure system stability and flexibility. The impact of the new system needs is different in each power system and depends on the size of the system (small or large), level of interconnection (strong or weak), {underlying flexibility of the existing portfolio, the available infrastructure of transmission (presence of bottlenecks), technological status (share of IBRs and DERs), regulatory policy etc. A tailor-made redesign of the AS markets must encourage eligible providers to supply the identified system needs, ensuring sufficient revenue and preserving investment signals for expansion in flexible capacity. 

Innovative technologies such as inverters, ESS, high voltage direct current (HVDC) grids, and information and communication technology (ICT) infrastructure, are paving the path toward a 100\% non-fossil future. Mathematical models are being developed to address the impacts of uncertainty and variability of VRE through forecasting techniques, and load modeling aims to capture the behavior of emerging loads. Financial incentives through price- or incentive-based programs are encouraging demand response and empowering consumers to unlock flexible resources. Jointly, innovative technologies, mathematical models, and demand response are key elements of a successful energy transition. These energy transition enablers direct the business plan and investments of AS providers. Market players can strategically place or manage innovative technologies at the transmission level to provide ASs by, for example, linking VRE generation with ESS, or interconnecting regions with HVDC grids. ICT infrastructure should be provided at the distribution level to adequately integrate DG, ESS, and EVs. Furthermore, ASs from heat, natural gas, hydrogen, and EVs may provide additional flexibility, promoting the coupling of the power system with other sectors.

The advances in US and EU power system operation practices and market designs serve as a reference point to other power systems due to the maturity of these wholesale markets. Accessing current technical barriers to improve system stability and market design flaws to mitigate inefficient economic signals allows practitioners to envision solutions and design modern AS markets under high shares of VRE. Also, investors can find opportunities to promote new business models. This paper provides a comprehensive review of more than two hundred papers (book chapters, reports, technical manuals) concerning the market design challenges in the US and EU, considering a power system transition toward a 100\% non-fossil power system future. Our paper contributes to the relevant literature as follows.

First, we contextualize how ASs have evolved and moved away from the synchronous-based power system paradigm. We clearly state the role of energy transition enablers in supporting ASs provision, addressing fundamental technological and modeling changes, and new demand response strategies. Second, we propose a holistic view of current US and EU ancillary service market designs through a systematic review of all US independent system operators (ISOs) and four relevant European power systems, considering their respective transmission system operators (TSOs). To make the proposed taxonomy easy to use, the AS types are conveniently summarized and categorized into frequency, non-frequency-related, and recently defined ASs. Third, we identify pivotal technological and market design barriers based on US and EU ancillary service market experiences. Subsequently, we propose potential solutions to enable secure and flexible power system operation under high VRE shares, to overcome market design inefficiencies that discourage providers from engaging in the AS markets. These potential solutions will help SOs and regulators redesign the existing AS markets.  

Table~\ref{tab:Review_comparison} compares our paper with other relevant papers addressing the ASs theme. No reviews were found, including existing and emerging ASs, systematically analyzing both US and European challenges and proposing potential solutions for modern AS markets under the transition towards a 100\% non-fossil future context.

\begin{table}[H]
\footnotesize
\centering
\caption{Comparative overview of some relevant review papers in the literature.}
\label{tab:Review_comparison}
\vspace{-3mm}
\footnotesize
\renewcommand{\arraystretch}{1.1} 
\linespread{0.9}\selectfont\centering
\begin{tabularx}{\linewidth}{*{1}{H{0.35}} *{1}{H{1.15}} *{1}{H{0.3}} *{1}{H{0.4}} *{1}{H{1.6}} *{1}{H{2.2}}}
\toprule
\footnotesize
Papers
& Power systems
& Existing AS
& Emerging AS
& Addressed challenges
& Observations
\\ \midrule 
\footnotesize
\cite{ReboursEcon2007}
& PJM,Europe,\newline New\,Zealand,\,Australia
& \checkmark
& \xmark
& AS market design
& VRE integration is not discussed
\\ 
\cite{Hirth2015}
& Europe
& \checkmark
& \xmark
& VRE integration in balancing markets
& Technologies and AS to integrate VRE are not discussed
\\ 
\cite{BanshwarBRIC2018}
& US,Brazil,Russia,\newline India,China
& \checkmark
& \xmark
& AS market after\,deregulation
& VRE is briefly mentioned
\\ 
\cite{Banshwar2017}
& Non-specific
& \checkmark
& \xmark
& AS from renewables
& DERs are briefly mentioned
\\
\cite{Pollitt2019}
& Non-specific
& \checkmark
& \xmark
& Competition in AS market
& Technical challenges are not discussed
\\
\cite{Jay2021}
& Non-specific
& \checkmark
& \xmark
& Reactive power AS market  
& It focuses only on a specific AS
\\
\cite{Fernandez2020}
& North\,America,\,Europe,\newline Australia,New\,Zealand
& \checkmark
& \checkmark
& Fast\,frequency\,control\,AS\,design
& It focuses only on a specific AS
\\
\cite{Sevdari2022}
& Non-specific
& \checkmark
& \checkmark
& AS from EV
& Other DERs and technologies are not discussed
\\
\cite{Delaney2020}
& Ireland
& \checkmark
& \checkmark
& New AS in Ireland
& The gaps to increase VRE share are briefly mentioned
\\
\cite{Rancilio2022}
& Europe
& \checkmark
& \checkmark
& Evolution of AS market
& Technical challenges under high VRE share are not discussed
\\
\cite{Oureilidis2020}
& Non-specific
& \checkmark
& \checkmark
& AS in distribution networks
& Inefficiencies in wholesale market design are not discussed
\\
\cite{Nordstroem2023}
& ERCOT and Europe
& \checkmark
& \checkmark
& System balancing under high VRE share
& It focuses only on system balancing
\\
Our Paper
& US, Europe
& \checkmark
& \checkmark
& \!Technical\,and\,market\,challenges for\,a\,100\%\,non-fossil\,future
& Potential technical solutions and market arrangements
\\
\bottomrule
\end{tabularx}
\end{table}

The authors in~\cite{ReboursEcon2007,Hirth2015,BanshwarBRIC2018,Banshwar2017,Pollitt2019,Jay2021} present relevant AS market design issues regarding the existing ASs as highlighted in Table~\ref{tab:Review_comparison}; however, technical or market design challenges under high shares of VRE are not discussed. Among the papers that include a comprehensive discussion about emerging ASs,~\cite{Fernandez2020} focuses specifically on fast response reserves and~\cite{Sevdari2022} in ASs provided by EVs. Paper~\cite{Delaney2020} only investigates the AS markets in Ireland/Northern Ireland. The authors in~\cite{Rancilio2022,Oureilidis2020} provide worldwide experiences, but no systematic comparison of AS markets and current practices is conducted. Although~\cite{Nordstroem2023} compares several power systems, the paper focuses only on power system balancing challenges. Therefore, our paper fills an existing gap in the literature by drawing a line on the limits of some current AS markets design to accommodate the transition towards a 100\% non-fossil power system and proposing potential redesign solutions.

\section{Ancillary Services}

The definition of ASs is intrinsically associated with a system operator's fundamental duty of ensuring reliable power system operation. Because procuring electrical energy and capacity does not guarantee system security, SOs must acquire a set of auxiliary (or ancillary) services from capable providers to satisfy secure power system operation~\cite{Kirschen2019}. Firstly, we detail the operating requirements that define conventional ASs designed from the perspective of a synchronous-based power system. Subsequently, we contextualize the ongoing energy transition, highlighting its enablers that may favor the provision of ASs. We propose a sufficient definition and taxonomy for ASs while introducing a number of new recently defined services. Moreover, we introduce and discuss the key fundamentals for the design of efficient AS markets.

\subsection{Conventional Ancillary Services}

Two main groups define the conventional provision of ASs. The first comprises services to maintain the active power balance in normal operation and after a contingency, which are referred to hereafter as \textit{frequency-related} ASs. The second includes services to ensure the reactive power balance and reserves to restore the power system after a significant contingency, which are referred to hereafter as \textit{non-frequency-related} ASs. 

\subsubsection{Frequency-Related Ancillary Services}

The inertial response (IR) and a hierarchical control scheme guarantee frequency equilibrium in steady-state operation. IR is an inherent reserve immediately released from synchronous machines after any active power imbalance between the generation and demand. This reserve is the stored rotational energy of the synchronous machines, which counteract rotational speed changes and is independent of the machine power output~\cite{Denholm2020}. An SG, equipped with a governor, can automatically change its active power when the frequency limit exceeds a defined dead band during regular or random deviations from the supply and demand, performing primary frequency control (PFC). In practice, if a tight tolerance band was imposed by SOs for nominal frequency under normal operating conditions, frequency fluctuations generally will not trigger PFC~\cite{Kirby2007,Andersson2012}. To maintain the frequency close to the nominal value, restore scheduled power flows between interconnected control areas, and reduce the area control error, secondary frequency control (SFC) is essential~\cite{Kirby2004,UCTE2004}. SFC is centralized and can be performed manually or automatically. Online generators with automatic generation control (AGC) can quickly change their active power output in response to rapid imbalances (second-to-second through minute-to-minute). In contrast, SO instructions or manual changes in dispatch can manage slower fluctuations (intra- and inter-hour)~\cite{Hirst1999}. Figure~\ref{fig:Control} shows a frequency excursion after a large under-frequency event and the activation sequence of the hierarchical control for a large power system.

\begin{figure}[!h]
    \centering
    \includegraphics[width=0.85\textwidth]{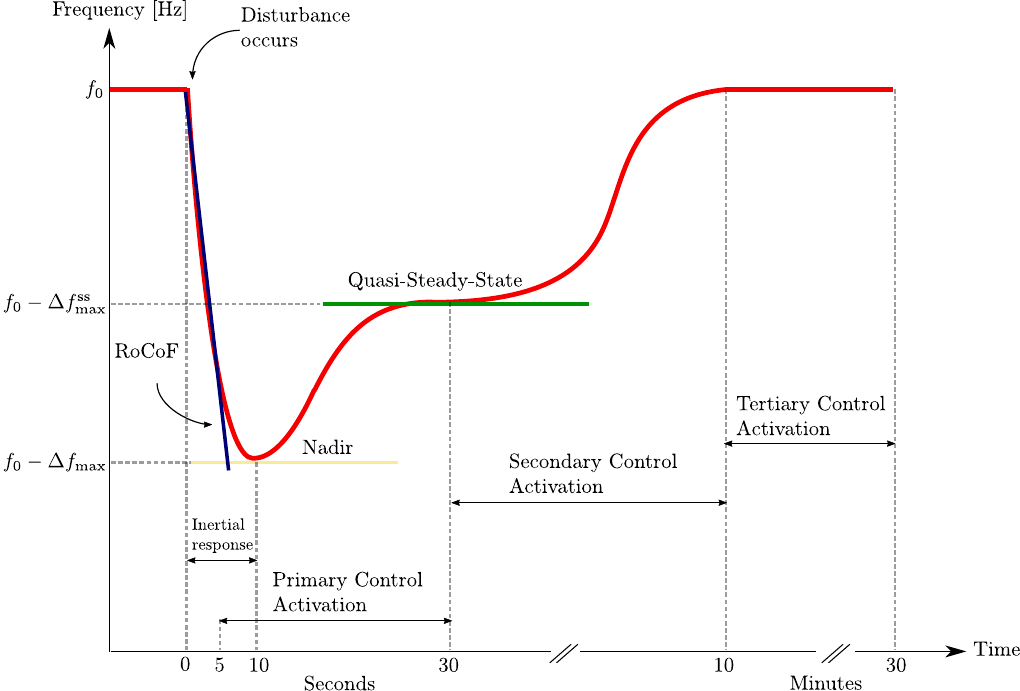}
    \caption{After the inertial response, a hierarchical control brings the frequency back to the nominal value.}
    \label{fig:Control}
\end{figure}

IR dampens the sudden frequency fall and extends the available response time until PFC acts, avoiding the disconnection of loads by under-frequency load-shedding (UFLS) schemes~\cite{Tielens2016,Ellison2012}. The large frequency variation experienced after a disturbance is outside the governor dead band. PFC captures the frequency drop and stabilizes the frequency within an acceptable range. The remaining frequency deviation from the nominal value is corrected through the SFC. Generators that were following the AGC signal under normal operating conditions can momentarily turn off AGC contributing to power capacity~\cite{Papalexopoulos2014}. Also, available capacity is manually activated from online and offline generating units~\cite{UCTE2004}. Deploying the reserve capacity of generators by dispatching it to generate electricity then results in a deficit of remaining reserves that must be replenished by rescheduling the generating units. Tertiary frequency control (TFC) restores the power system to the pre-contingency status, preparing it for the next possible contingency~\cite{Ela2011}. 

Three important operational metrics to limit the system frequency are illustrated in Fig.~\ref{fig:Control}. Immediately following the contingency event, the derivative of frequency to time defines the maximum rate of change of frequency (RoCoF). RoCoF is an operational metric that indicates how fast the frequency changes. The nadir occurs at the maximum frequency deviation ($\Delta f_{\rm{max}}$) point~\cite{Rezkalla2018}, while the maximum allowed frequency deviation ($\Delta f_{\rm{max}}^{{\rm{ss}}}$) sets the quasi-steady-state frequency, that is notionally the frequency reached after inertial and primary response but before secondary response activation, as shown in Fig.~\ref{fig:Control}. These operational metrics help SOs define the necessary reserves and set protection schemes~\cite{Kirby2004,Jorgenson2019}.

\subsubsection{Non-Frequency-Related Ancillary Services}

Besides the system frequency, reactive power imbalances can be regulated by monitoring voltage variations. To operate seamlessly, electrical equipment needs to function within a narrow voltage band; otherwise, malfunctions and damage can occur. The supply of reactive power results in the consumption of generation and transmission resources. Since reactive power losses increase with distance, voltage control is location-constrained. Therefore, static and dynamic devices are installed at key buses in the power system to provide this AS~\cite{Kirby1997,FERC2005}. Static devices, such as capacitors and reactors, help to regulate steady-state voltages. Dynamic devices can control the voltage output in response to voltage changes, such as flexible AC transmission system (FACTS) devices. The latter includes static VAR compensators (SVC) and static synchronous compensators (STATCOM). Notice that a synchronous generator continuously adjusts its reactive power to perform systemic voltage control. Also, synchronous condensers (SC) may be installed for an improved reactive compensation~\cite{Fusco2007,FERC2005}.

In the event of a blackout, power system restoration must be initiated as quickly as possible to minimize technical and economic losses. This task involves complex coordinated steps, commencing with the restart of the generators. In this case, the necessary steps include restarting appropriate resources rapidly without an external power supply, energizing transmission lines, and restarting other available generators~\cite{Kirby2007}. Generally, relatively small power plants, such as hydroelectric power plants, pumped hydropower storage (PHS), and combustion turbines, with a battery or diesel generator to feed the auxiliaries of the main generator, start the restoration process~\cite{Gracia2019}. 

\subsection{Enabling the Energy Transition}\label{sec:AS:enabling}

The transition to a 100\% non-fossil power system is a challenging path, where the operating requirements of the power system must be attained under high shares of VRE. Existing synchronous resources allow the introduction of new technologies maintaining system stability. Innovative technologies are decarbonizing the power system and boosting the energy transition. Mathematical models should be created or adjusted to include a more accurate representation of the new system needs, and new demand response strategies should facilitate power system decentralization. This section highlights the main enablers of the energy transition capable of supporting the provision of ASs.

\subsubsection{Existing Synchronous Resources}

The evolution to an inverter-based paradigm is notably sustained by continuous experimentation in a changing power system. The capabilities of the existing synchronous resources allow a gradual introduction of new technologies to avoid system instabilities. For instance, Ireland/Northern Ireland is imposing a 75\% limit for non-synchronous instantaneous generation penetration to maintain secure operation~\cite{EirGridSNSP2023}. A minimum number of online synchronous units may be necessary to guarantee stable operation, considering the VRE-driven displacement of SGs. Synchronous condensers can mitigate the reduced system inertia and synchronizing torque levels, enhancing the dynamic voltage support and fault levels provision~\cite{EirGridmit2021}. Considering adequate financial incentives, coal-fired power plant owners can also retrofit their units to improve operational flexibility, such as reducing their stable minimum power output to expand the operational range~\cite{IESR2020}. The unlocked flexibility of coal-fired plants is helpful in avoiding the curtailment of VRE generation if network bottlenecks constrain the power transfer.

\subsubsection{Innovative Technologies}

Emerging technologies which transform the traditional synchronous-based paradigm and promote increasing shares of VRE can be named innovative technologies. Next, we detail advances in these technologies, indicating promising developments to ensure secure and flexible operation.

\paragraph{Power Electronics and Control}

Power electronics are essential in converting and controlling raw VRE and deploying ESS, which are useful for managing load fluctuations. However, the electronic interface decouples variable speed wind turbines (VSWT) (particularly full converter type) and photovoltaic (PV) generators from the grid, preventing the natural transient response under a contingency. Consequently, they do not inherently provide IR, which leads to lower system inertia if conventional generators are displaced in the economic dispatch, such as in high instantaneous VRE share conditions. The reduction in system inertia leads to an increase in RoCoF and a lower frequency nadir if the largest contingency remains the same, thus requiring that other generators respond in a shorter time frame~\cite{Kroposki2017,Milano2018}.

As of the writing of this paper, most IBRs are coupled to grid-following (GFL) inverters, acting as a controlled current source. A current control loop quickly changes the current output based on the angular reference from the phase-lock loop (PLL) control. The PLL estimates the instantaneous voltage phase angle by measuring the terminal voltage phasor of the inverter. Several control techniques, based on the frequency measurement, have been considered to enable a frequency response when GFL inverters are adopted~\cite{Dreidy2017,Fernandez-Guillamon2019}. The \textit{hidden inertia} technique comprises a supplementary control that allows a VSWT to respond rapidly to frequency changes. After a disturbance, the power output of the wind turbine can be increased based on the frequency deviation, which slows down the turbine and enables the release of the hidden rotational energy from the rotating mass~\cite{Morren2006,Muljadi2012}. Alternatively, \textit{fast power reserve} aims to increase the power output within a constant percentage~\cite{Ullah2008} or range (e.g. 5-10\% as in~\cite{Gevorgian2015}) of the nominal wind turbine power for a defined wind speed range. The overproduction period helps to arrest the RoCoF; however, it is followed by an underproduction period due to operation below the maximum power point. Deloading of the VSWT or PV generation provides a reserve margin for PFC activation. A governor-like behavior is emulated through \textit{droop control} programmed in the inverter to respond to frequency deviations by changing the active power proportionally, improving the frequency nadir~\cite{Wu2018,Rajan2021}.   

In addition to a frequency response, the power electronic interface can assist VSWT and PV panels in voltage control. Under normal operating conditions, a VSWT supports reactive power by controlling the voltage of a specific bus or setting a fixed power factor. However, these control strategies can be insufficient if the power plant is placed in a weak grid to maximize the use of wind or solar resources. Also, system stability should be ensured in transient periods, and thus, additional control strategies and FACTS devices, such as STATCOM, and SVC, can ensure the voltage ride-through requirements imposed in grid codes~\cite{Vittal2012,Shah2015}. Solar PV panels inherently produce DC power and cannot deliver reactive power. Nevertheless, modern inverters can absorb or inject reactive power from the grid, performing voltage control~\cite{Sarkar2018}. In~\cite{Varma2015}, PV inverters are programmed to act as a STATCOM, during the night and day to avoid voltage instabilities. Voltage ride-through capabilities are accessed in VSWTs through improved control techniques or connection of external FACT devices~\cite{Hiremath2020}.  

Inverters provide a low short circuit fault contribution, reducing system strength. Additionally, an inherent delay in processing the signal from the PLL control is inevitable in GFL inverters preventing an immediate response to disturbances. Under extremely high instantaneous VRE share, system stability may be seriously threatened. Grid-forming (GFM) inverters operate as voltage sources, imposing a constant voltage phasor without the need for a PLL~\cite{Lin2020,NERC2021GFM}. Therefore, an essentially instantaneous response can be obtained. Using some short-term ESS (batteries or supercapacitors) or sufficient headroom from the input energy source (wind or solar), and a modified control strategy, such as the so-called \textit{virtual synchronous machine} (VSM), the dynamic behavior of a synchronous machine under disturbances can be emulated~\cite{Matevosyan2019GFM,Migrate2018}. Several control techniques are summarized in~\cite{Bevrani2014} and~\cite{Tamrakar2017}. In addition, GFM inverters allow PV and wind plants to energize their own sites. A coordinated process can create smaller and distributed power islands in the distribution system that will further energize the transmission lines~\cite{NGnblack2019}.

\paragraph{Energy Storage Systems}

An ESS can shift energy and power in time and they are crucial for decarbonizing the power system since they act as an energy buffer smoothing variable generation. They can provide short- and long-term capacity, enhancing system flexibility to alleviate the peak load, deferring grid investments, and providing frequency and voltage control. Available options include mature technologies, such as pumped hydropower storage (PHS), more recent developments, such as battery energy storage systems (BESS) and flywheels, and newer solutions, such as supercapacitors and hydrogen storage. A hybrid system, including two or more technologies, is also possible~\cite{Eyer2010,Molina2017}.

A PHS is a versatile technology to provides frequency-related AS in generating mode. In pumping mode, fixed-speed PHS can operate their SG as a synchronous condenser to increase the inertia of the load. Also, variable-speed PHS coupled by an inverter interface can contribute to a fast response against frequency deviations after a disturbance~\cite{IRENApump2020,Nag2022}. Using batteries in EVs make them economical for power system applications. BESS can quickly respond to changes in system frequency due to the absence of moving components. In~\cite{Zhang2016ying}, the authors formulate an optimal state of charge strategy for PFC provision. Similarly, SFC using BESS is analyzed in~\cite{Canevese2017} to evaluate the impact of continuous cycling on battery aging. BESS are also helpful to enhance short-term flexibility due to their high ramp capability~\cite{Hu2018} and can contribute to system restoration if suitably located~\cite{Yao2020}. Flywheels are well-suited to follow small frequency imbalances~\cite{Lazarewicz2004}, and the short-term storage of supercapacitors can assist in inertia emulation control~\cite{Xiong2018}. Additionally, long-term hydrogen storage can potentially mitigate seasonal fluctuations reducing the curtailment of variable generation, as presented in~\cite{Elberry2021}. 

\paragraph{High Voltage Direct Current Grids}

HVDC transmission lines provide power transfer for long lines, interconnecting systems asynchronously and helping to integrate renewable energy by enabling energy balancing over a wider area}. Line-commutated converter (LCC) HVDC links have a high power transfer rating, but their inability to provide an AC voltage from the DC side precludes black start operation. Also, LCC HVDC cannot support voltage control. Voltage source converter (VSC) HVDC links have black start capability and can contribute to voltage control due to the independent controllability of active and reactive power~\cite{Kaushal2019,Watson2020}. 

Additionally, LCC and VSC HVDC technologies can contribute to frequency response. Considering LCC HVDC links, the authors in~\cite{Kwon2020} propose integrating the converters with feedback loops to emulate inertia and provide PFC. Regarding VSC HVDC technology,~\cite{Shen2021} proposes a scheme capable of autonomously adjusting the emulated inertia constant according to the grid frequency deviations. Detailed frequency control strategies are presented in~\cite{Lin2021}. The focus is on VSC HVDC transmission lines connected to wind farms.

\paragraph{Information and Communication Technology}

Appropriate ICT infrastructure is essential to monitor, control, and manage resources at different voltage levels. In distribution systems, households can interconnect sensors and controllers through an internet of things (IoT) infrastructure to facilitate the automation of residential appliances and optimize their electricity consumption~\cite{Wen2015,Sajadi2018}. Digital meters can enable two-way communication between utilities and consumers, improving the visibility of rooftop PV systems, BESS, and EVs across the grid, allowing consumers to trade their flexibility. 

System observability can be improved using wide-area measurements from phasor measurement units (PMUs), enhancing system security. Time-synchronized phasor (voltage and current) measurements enable real-time frequency and voltage monitoring, fault and oscillation detection, allowing SOs to adopt corrective actions faster~\cite{Joshi2021}. Also, inertia estimation is essential under high shares of VRE. Model-based methods derived from the swing equation and real-time estimation from synchrophasor data 
are found in the literature~\cite{Tan2022}.

\subsubsection{Mathematical Models for Power System Operation and Stability }

More accurate wind and solar power forecasts, and improved representation of modern loads arising from dispersed resources at low-voltage distribution systems, illustrate how mathematical models can be applied in power system planning and operation to enhance AS provision. In this section, we show why some models are outdated, considering the transition to a 100\% non-fossil future and the benefits of their improvements.  

\paragraph{Wind and Solar Power Forecasting Models}

Forecasting models have been developed to support improved decisions from SOs and market participants due to the stochastic nature of wind and solar. Methodologies can be divided into deterministic and probabilistic methods. Deterministic methods provide a single series of expected values and are widely used in power systems by SOs to guarantee sufficient reserve. In electricity markets, the deterministic forecast guides VRE producers to find suitable trading strategies. Significant advantages are simplicity, compatibility with existing operator tools, straightforward evaluation, and fast use and reproduction. Deterministic methods are classified as physical or statistical models. Numerical weather prediction is a physical model based on meteorological data suitable for long-term horizons (a day to a week ahead). Time series-based and artificial intelligence (AI) models (or a hybrid approach) are statistical models adequate for short-term (minutes to hours ahead) forecasts~\cite{Soman2010,Mohammadi2020}. 

Probabilistic methods provide a confidence interval of the expected values, resulting in uncertainty estimation. A range of possible outcomes is an improved solution compared to the deterministic point forecast. Analysis of multiple scenarios potentially optimizes the SO reserve procurement and the decision-making process of VRE suppliers~\cite{Bazionis2021}. Probabilistic methods can be parametric or non-parametric. The former is based on a known probability density function. In contrast, no assumptions are made in non-parametric models, and a tailor-made probability density function is empirically determined~\cite{Van2018}. 

\paragraph{Load Modeling}

Traditional static and passive load models, such as the \textit{constant impedance, current, power} model (ZIP), and \textit{exponential model}, have been extensively adopted in power system analysis~\cite{Kundur1994}. However, changes in load profile as the increasing participation of drive-controlled induction motors, for instance, in air conditioning systems, and the proliferation of DERs, make the traditional load models less suitable to represent load behavior. Dynamic load models, such as the \textit{induction motor} (IM) model, incorporate differential equations derived from the load equivalent circuit to describe the active and reactive power response in time, and is particularly helpful for angular and transient stability studies~\cite{NERCload2016}. Additionally, static and dynamic load model components can be aggregated to create a \textit{composite load} model, such as (ZIP and IM), which tend to be more accurate than the individual models~\cite{Renmu2006}. 

After choosing a load model structure (static, dynamic, or composite), a key step in load modeling is the identification of load parameters to validate the model. Existing methodologies are divided into component- and measurement-based approaches. The former relies on information about electricity consumption, that is, the load composition, and is available in commercial tools. Customers with similar load composition are aggregated in classes of loads, typically residential, commercial, industrial, agricultural, and others. The latter aims to fit a load model that emulates the behavior of the aggregate load model recorded from disturbance measurements of PMUs and digital meters~\cite{Milanovic2014, Arif2018, IEEE2022}. The component-based approach does not require field measurements, saving costs on installing measurement devices. However, gathering detailed information about load composition is difficult. Load profile may change if consumers engage in demand response programs and start to follow instructed price signals. Also, the load composition varies seasonally and in the short-term (daily and weekly). In contrast, measurements are valuable for SOs because they indicate the operating conditions. The main advantage of the measurement-based approach is retrieving the power system dynamic response. Nonetheless, this approach cannot provide generic load models since data is collected at specific locations and depends on the occurrence of disturbances. Combining model- and data-driven approaches in a hybrid solution is a potential alternative~\cite{Milanovic2014, Arif2018, IEEE2022}. 

Emerging modern loads at low voltage levels, such as EVs, solar rooftop PV systems with or without batteries, and heating and cooling electrical loads, are typically hidden behind a power electronic interface, resulting in a non-linear relation between voltage and current. The complexity of the current distribution system makes the use of detailed mathematical models for load modeling computationally impractical. A dynamic equivalent model of active resources can be obtained using disturbance measurements and some system identification method, such as artificial neural networks. By aggregating the behavior of a large number of resources with different technologies, SOs simplify the dynamic analysis of active distribution networks~\cite{Resende2013}. If SOs can adequately quantify the new system needs arising from the distribution network, new grid code requirements and the need for new ASs may be well addressed.

\subsubsection{Demand Response}\label{sec:AS:demand_response}

Demand-side participation is a powerful strategy to engage large and small players, such as industrial, commercial, and residential loads, in AS provision. Next, we discuss three potential strategies to increase power system flexibility and competition in the wholesale electricity market.  

\paragraph{Sector Coupling}

Coupling power systems with heat, gas, hydrogen, and transportation sectors allows SOs to procure additional flexible resources. For instance, power-to-heat includes thermal loads (building heating and cooling, water heating, refrigeration, and freezing), water pumping, air compression, and loads with associated storage processes~\cite{Heffner2007}. These resources can assist in frequency control, as shown in~\cite{Corsetti2020}. The authors simulate a real district-heating plant to regulate frequency imbalances, stressing the importance of multi-energy systems. Hydrogen is an energy vector particularly helpful in storing and transporting renewable energy produced in power systems. The production of green hydrogen by electrolysis using excess power from VRE or hydropower plants is a power-to-hydrogen application that can enhance power system flexibility and contribute to the decarbonization process. Also, electrolyzers can act as a controllable load modulating the hydrogen production and quickly responding to frequency deviations, as presented in~\cite{Dozein2021}, benefiting from demand response price signals. The vehicle-to-grid strategy involves a bidirectional control that allows electricity stored in vehicle batteries to be pushed back into the grid. These mobile batteries could allow EVs to respond rapidly to frequency deviations. In~\cite{Kempton2008}, a single EV is used as a proof of concept to track frequency and store energy. Capacity payments should incentivize EV owners to maintain their vehicles parked and recover additional battery life costs due to increased cycling and round-trip energy losses.

\paragraph{DER Aggregation}

\begin{figure}[!h]
    \centering
    \includegraphics[width=\textwidth]{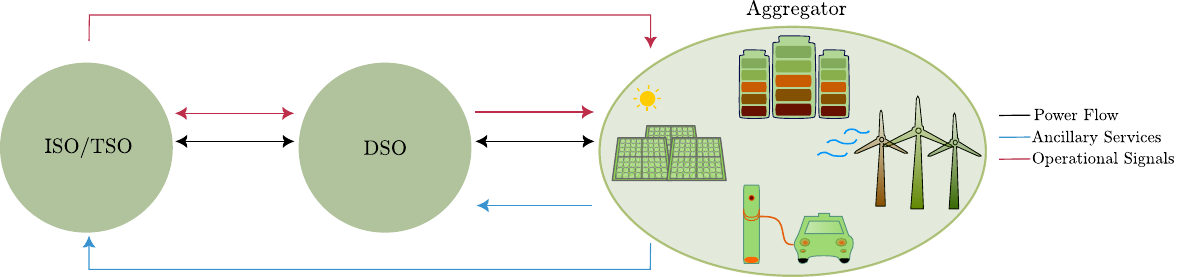}
    \caption{SO-DSO-aggregator coordination relies on the exchange of information between agents to provide electricity and services (based on~\cite{Birk2017}).}
    \label{fig:Aggregator}
\end{figure}

Resources in the distribution system are dispersed and have smaller capacity compared to typical transmission-connected resources, but are potential providers of ASs. DERs can be aggregated to form a virtual power plant (VPP) or an energy community (EC). In contrast to a VPP, an EC primarily focuses on social, economic, and environmental benefits, rather than financial profits~\cite{EC2020ec}. An aggregator is a third-party company that coordinates several DERs and acts as an intermediary between DER owners, the SO, and the DSO. The aggregation of DERs is changing the traditional centralized generation paradigm, empowering consumers to become producers (prosumers), and service suppliers in a decentralized fashion. Centrally coordinating individual DERs through bidirectional communication is impractical for SOs. Instead, an aggregator can interface thousands of DERs and receive operational signals with instructions from DSOs and SOs~\cite{Birk2017}, as shown in Fig.~\ref{fig:Aggregator}. 

Aggregators could be allowed to provide services locally for the DSO or directly for the SO. To avoid the risk of double counting resources or for SO and DSO to be operating at cross-purposes, close coordination and a clear definition of responsibilities between SO and DSO are fundamental to ensure that transmission and distribution network constraints will be respected while aggregators
can compete with large players~\cite{ESIG2022}. 

Considering a VRE-based VPP, the authors in~\cite{Camal2022} propose a methodology to improve the forecasting performance of aggregate wind, solar, and hydroelectric power on extreme quantiles to reduce the risk of not providing ASs. The power production obtained is used to offer reserve capacity to follow downward movements of the system frequency. In~\cite{Zhang2016tian}, a hierarchical energy management system is proposed to optimize the operation of an aggregated BESS supplying electricity and frequency regulation. The authors consider the performance of the battery to follow the regulation signal and the coordination of two different battery types. EV aggregation for frequency regulation provision is discussed in~\cite{Meng2015}. The comfort level of EV owners is reduced if frequency regulation capability increases, irrespective of the charging strategy adopted, but discharging EV batteries during load valley optimizes the frequency regulation capability throughout the day. Two-stage stochastic programming is used to model an energy community considering the active and reactive power provision from DERs to DSOs in~\cite{Garcia2022}. A collaborating scheme rewards reactive power supply and reduces the total community cost.  

\paragraph{Data Centers}

The growing internet use to enable numerous applications of digital technologies relies on electricity-intensive data centers. In 2022, global electricity consumption from data centers was estimated at around 1-1.3\% of annual electricity demand\footnote{Data centers for cryptocurrency mining were not included in this estimation. They correspond to an additional 0.4\% of global annual electricity demand~\cite{IEA2023}.}~\cite{IEA2023}. Since digital transformation tends to increase, it is valuable to identify the potential interplay of data centers in the power system. Firstly, data centers are an important flexible load. To maintain continuous operation during a power outage, uninterruptible power supply (UPS) systems, typically formed by redundant BESS, are installed to provide the necessary backup. Nevertheless, the redundancy of the backup system oversizes the capacity of the batteries, which are rarely used due to a stable power supply. Thus, there is an opportunity for a revenue stream, if data centers partially operate their workload using their flexibility from idle energy storage capacity, and financial incentives are provided~\cite{SmartEn2021}. Delay-tolerant workloads can be shifted in time, and workloads also can be routed to other data centers dispersed geographically~\cite{Wierman2014}. Such temporal and spatial load management allows data centers to procure electricity when and where it is greener and cheaper, contributing to power system balancing and decarbonization.

Secondly, several pilot projects have demonstrated that UPS can provide frequency response. Fast-acting reserve provision from a UPS has been tested in Ireland and the Nordic power system. Small data centers, with limited UPS storage capacity, would face barriers to competing as an AS provider. Therefore, the trials also considered the participation of the data center in a VPP. In US, a UPS has been considered to demonstrate frequency regulation following PJM signals~\cite{Paananen2021}. Thirdly, electricity consumed by data centers generates heat as a byproduct, which can be funneled into a district heating network to be reused for residential and commercial buildings. Existing initiatives are found in Ireland, and the Nordic power system~\cite{OShea2019,Data2020}. However, the low temperature of the waste heat from data centers requires additional heat pumps to raise the temperature, which consumes electricity and increases total costs~\cite{Wahlroos2018}.   

\subsection{Emerging Ancillary Services}

To handle the impacts of reduced inertia levels, some immediate solutions include managing the potential of conventional technologies. Traditionally, synchronous inertia and PFC are byproducts of SG operation. In power systems with high shares of VRE, the operation of some conventional SGs can be uneconomical. However, these units are critical to guarantee a minimum inertia level and frequency stability. \textit{Synchronous inertial response} (SIR) should be explicitly defined as a new AS to encourage synchronous resources to reduce their minimum generation level and allow additional units to remain online, contributing to their rotational energy. SIR was introduced in EirGrid/SONI (Ireland/Northern Ireland)~\cite{Delaney2020} and was discussed in the ERCOT AS market redesign process~\cite{Matevosyan2015}. \textit{Primary frequency response} (PFR) is the rapid response to changes in frequency. SGs provide PFR capability through the local and automatic electromechanical control of the turbine governor. Alternatively, PFR can be delivered using controllable loads with a governor-like response. PFR is a well-developed AS in Europe and is being introduced in ERCOT as an explicit AS, moving away from the obligatory requirement traditionally adopted in US. As shown in Section~\ref{sec:AS:enabling}, IBRs are evolving and contributing to system security. Creating new ASs to include technology-agnostic solutions and removing market design barriers of existing ASs, which favor SG-based solutions, is essential to incentivize capable resources toward a competitive electricity market. Non-synchronous resources interfaced by GFM inverters with a modified control strategy and some energy buffer to act like the rotational energy of synchronous resources can emulate the inertial response of synchronous machines, virtually providing inertia. Currently, no SO explicitly defines a \textit{virtual inertial response} (VIR) service to fit the capabilities enabled by GFM inverters. Non-synchronous resources can respond to very fast changes in frequency independently of the inverter type. In this sense, \textit{fast frequency response} (FFR) is another newly defined AS introduced by several SOs, comprising a subset of PFR, to support a faster acting capability than the traditional governor response. FFR capability can be obtained by shedding large industrial interruptible loads triggered by under-frequency relays or using the active power capability from VRE, HVDC links, or BESS.

Wind and solar generation naturally introduce variability and uncertainty due to weather dependence. Deviations in the VRE generation forecast and net load can occur, resulting in significant power imbalances and, consequently, increased ramps. Thus, existing resources should be encouraged to enhance their flexibility, and new units should have financial incentives to promote a flexible operation. \textit{Flexible ramping capability} (FRC) comprises the ramping capability from flexible resources, online or offline, capable of quickly ramping, following future movements of the net load. This newly distinguished AS is procured in CAISO~\cite{CAISOFRP2018}, MISO (Midwest US)~\cite{Navid2013}, SPP (central Southern US)~\cite{SPPramp2022}, and EirGrid/SONI~\cite{Delaney2020}.

The displacement of SGs can reduce the dynamic reactive capability. In EirGrid/SONI, the \textit{dynamic reactive response} (DRR) incentivizes resources to provide a fast reactive response after an event. The \textit{fast post-fault active power recovery} (FPFAPR) is another newly defined AS that rewards wind generators capable of quickly recovering their active power output after a large voltage dip that impacts frequency stability~\cite{EirGridproposal2013}.

\subsection{Taxonomy of Ancillary Services}

\begin{figure}[!h]
    \centering
    \tikzset{box/.style={draw, rectangle, rounded corners, thick, node distance=1em, text centered, minimum height=3em}}
    \tikzset{whtblock/.style={rectangle, draw, fill=white!20, text width=18mm, text centered, minimum height=4em},
    shdblock/.style={rectangle, draw, fill=lightgray!50, text width=18mm, text centered, minimum height=4em},
    header/.style={rectangle, draw, fill=white!20, text width=18mm, text centered,font=\fontsize{7pt}{7pt}\selectfont},  
    container/.style={draw, rectangle,dashed,inner sep=0.15cm, rounded corners,fill=yellow!20,minimum height=2.6cm}}
    
    \begin{tikzpicture}[node distance = 0.2cm, auto, every node/.style={font=\fontsize{7.5pt}{7.5pt}\selectfont}]
  
    \node [shdblock] (SyncIn) {Synchronous\\[0.2ex] Inertial\\[0.2ex] Response\\[0.2ex] \textcolor{blue}{(continuous)}};   

    
    \node [shdblock, below=of SyncIn, node distance=2.5cm] (VirIn) {Virtual\\[0.2ex] Inertial\\[0.2ex] Response\\[0.2ex] \textcolor{blue}{(event-driven)}};


\node [shdblock, right=2.5cm of SyncIn.center, anchor=center] (PFR) {Primary\\[0.2ex] Frequency\\[0.2ex] Response\\\textcolor{blue}{(primarily\\[0.2ex] event-driven)}};


\node [shdblock, below=of PFR] (FFR) {Fast\\[0.2ex] Frequency\\[0.2ex] Response\\[0.2ex]\textcolor{blue}{(event-driven)}};


\node [whtblock, right=2.5cm of PFR.center, anchor=center] (FR) {Frequency\\[0.2ex] Regulation\\[0.2ex] (up/down)\\[0.2ex] \textcolor{blue}{(continuous)}};   


\node [whtblock, below= of FR] (SR) {Spinning\\[0.2ex] Reserve\\[0.2ex]\textcolor{blue}{(event-driven)}};


\node [whtblock, below=of SR] (NSR) {Non-Spinning\\[0.2ex] Reserve\\[0.2ex]\textcolor{blue}{(event-driven)}};


\node [whtblock, right=2.5cm of FR.center, anchor=center] (RR) {Replacement\\[0.2ex] Reserve\\[0.2ex]\textcolor{blue}{(event-driven)}};


\node [shdblock, right=2.5cm of RR.center, anchor=center] (FRC) {Flexible\\[0.2ex] Ramping\\[0.2ex] Capability\\[0.2ex] (up/down)\\[0.2ex]\textcolor{blue}{(continuous)}};


\node [whtblock, right=2.5cm of FRC.center, anchor=center] (SSRR) {Steady-State\\[0.2ex]Reactive\\[0.2ex] Response\\[0.2ex]\textcolor{blue}{(continuous)}};


\node [shdblock, below=of SSRR] (DRR) {Dynamic\\[0.2ex] Reactive\\[0.2ex]Response\\[0.2ex]\textcolor{blue}{(event-driven)}};



\node [shdblock, below=of DRR] (APR) {Fast Post-\\[0.2ex]Fault Active\\[0.2ex] Power\,Recovery\\[0.2ex]\textcolor{blue}{(event-driven)}};


\node [whtblock, right=2.5cm of SSRR.center, anchor=center] (BSC) {Black-Start Capability\\[0.2ex]\textcolor{blue}{(event-driven)}};


    \begin{scope}[on background layer]
        \coordinate (aux1) at ([yshift=5mm]SyncIn.north);
        \node [container,fit=(aux1) (VirIn)] (IR) {};
        \node at (IR.north) [header] {\textbf{Inertial\\[0.2ex]Response}};
  
        \coordinate (aux2) at ([yshift=5mm]PFR.north);
         \node[container, fit=(aux2) (FFR), right=2.5cm of IR.center, anchor=center] (PFC) {};
        \node at (PFC.north) [header] {\textbf{Primary\\[0.2ex]Frequency\\[0.2ex] Control}};
  
  
     \coordinate (aux3) at ([yshift=5mm]FR.north);
      \node[container, fit=(aux3) (SR) (NSR), right=2.5cm of PFC.north, anchor=north] (SFC) {};
        \node at (SFC.north) [header] {\textbf{Secondary\\[0.2ex]Frequency\\[0.2ex] Control}};

    \coordinate (aux4) at ([yshift=5mm]RR.north);
        \node[container, fit=(aux4) (RR), right=2.5cm of SFC.north, anchor=north] (Flex) {};
         \node at (Flex.north) [header] {\textbf{Tertiary\\[0.2ex]Frequency\\[0.2ex] Control}};
  
  
    \coordinate (aux5) at ([yshift=5mm]FRC.north);
     \node[container, fit=(aux5) (FRC), right=2.5cm of Flex.north, anchor=north] (TFC) {};
        \node at (TFC.north) [header] {\textbf{Economic\\[0.2ex]Dispatch}};

   
    \coordinate (aux6) at ([yshift=5mm] SSRR.north);
        \node[container, fit=(aux6) (DRR) (APR), right=2.5cm of TFC.north, anchor=north] (VC) {};
        \node at (VC.north) [header] {\textbf{Voltage\\[0.2ex] Control}};
  
  
  \coordinate (aux7) at ([yshift=5mm]BSC.north);
        \node[container, fit=(aux7) (BSC), right=2.5cm of VC.north, anchor=north] (ST) {};
        \node at (ST.north) [header] {\textbf{System\\[0.2ex] Restoration}};
  
\end{scope}
\end{tikzpicture}
     \caption{Proposed taxonomy of ASs in power systems with a significant share of VRE. Shaded rectangles emphasize the emerging ASs.}
    \label{fig:Taxonomy}
\end{figure}
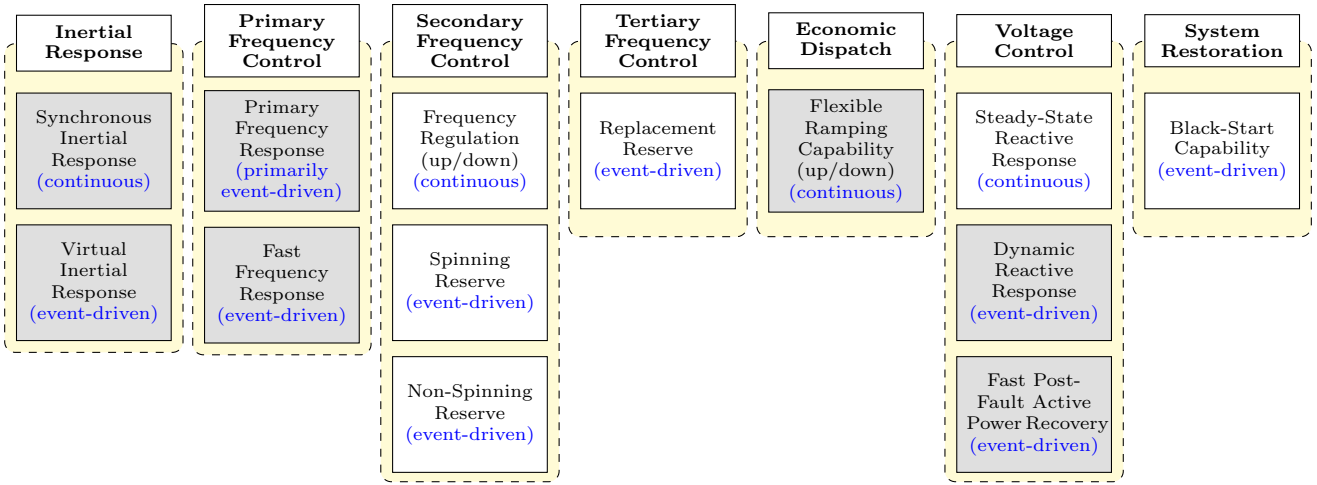

The absence of a standard nomenclature and definition between SOs can lead to misunderstandings by researchers and industrial practitioners, used to typical textbook or regional SO nomenclatures. We propose a comprehensive definition for each AS, highlighting potential providers, as shown in Table~\ref{tab:AS_definition}. Figure~\ref{fig:Taxonomy} links the conventional and recently defined ASs according to their role in system security. The proposed taxonomy is not exhaustive, noting that other potential new ASs are proposed in the literature~\cite{Oureilidis2020}.

\begin{table}[!h]
    \footnotesize
    \centering
    \caption{Proposed definition and potential providers of conventional and emerging ASs.}
    \vspace{-2mm}
    \label{tab:AS_definition}
    \begin{tabularx}{\linewidth}{ *{1}{H{0.71}} *{1}{H{0.23}}  *{1}{H{1.9}} *{1}{H{1.1}}  } 
    \hline\\[-2.5ex]
    \textbf{Ancillary Service}&  & \textbf{Definition}& \textbf{Providers} \\[0.5ex]\hline\\[-2.5ex]
    \textit{Synchronous \newline Inertial Response} & SIR &  Inherent and uncontrolled response from synchronous \newline resources& SG,\,SC,\,PHS,\,synchronous\,loads\\[3.5ex]
    \textit{Virtual Inertial \newline Response}& 
    VIR& Emulation of synchronous inertia from non-synchronous resources responding proportionally to RoCoF& BESS,\,EVs,\,supercapacitor,\,wind\newline or solar headroom, energy\,stored in\,HVDC\,links\\[6.5ex]
    \textit{Primary\,Frequency\newline Response}&PFR& Active power capability to quickly respond to frequency deviations by increasing generation or reducing demand&SG and controllable loads\\[3.5ex]
    \textit{Fast Frequency\newline Response}&FFR& Faster primary frequency response capability&VRE, BESS, EVs, HVDC\,links,\newline interruptible loads\\[3.5ex]     \textit{Frequency\,Regulation}&FR&Centralized increase or decrease in active power provided by resources with AGC to match the nominal frequency at all times under normal operation&SG,\,VRE,\,BESS,\,flywheels,\,control-lable\,loads,\,DER\,aggregation\\[6.5ex] 
     \textit{Spinning Reserve}&SR&Active power capacity  synchronized to the grid, which can be deployed immediately under a contingency condition&\multirow{7}{*}{\parbox{\linewidth}{SG,\,VRE,\,BESS,\,DER aggre-gation,
     electrolyzer}}\\[6.5ex] 
     \textit{Non-Spinning\newline Reserve}&NSR&Active power capacity that, although non-synchronized to the grid, can be readily available within a few minutes under a contingency condition\\[6.5ex] 
     \textit{Replacement\,Reserve}&RR&Active power capacity necessary to restore the levels of SR and NSR to their pre-contingency status\\[3.5ex] 
     \textit{Flexible Ramping\newline Capability}&FRC&Ramping capability (increase or decrease in active power), provided by flexible resources, online or offline, to follow future movements of net load&VRE,\,BESS,\,SG\,(hydro,\,gas-fired), DER aggregation\\[6.5ex]
     \textit{Steady-State\newline Reactive Response}&SSRR&Reactive power capacity to maintain the voltage within narrow limits under normal operation& SG,SC,VRE,SVC,STATCOM,\newline DERs,\,capacitors,\,reactors
     \\[3.5ex]
    \textit{Dynamic Reactive\newline Response}&DRR& Fast reactive power response used during disturbances to recover the voltage to its nominal value& SG,\,SC,\,VRE,\,SVC,\,STATCOM,\newline DERs\\[3.5ex]
    \textit{Fast\,Post-Fault\newline Active\,Power\,Recovery}&FPFAPR& Fast recovery of inverter-based generation active power output after a disturbance.& VRE\\[3.5ex]
     \textit{Black Start\newline Capability}&BSC&Active and reactive power capacity responsible for starting the system restoration after a black-out& SG, BESS, PHS\\
     \hline
\end{tabularx}
\end{table}

Depending on the system, some services are implemented as mandatory capabilities within the grid codes, rather than explicit ASs procured and remunerated by SOs. Some examples are capabilities for power quality improvement (power smoothing, harmonic mitigation, and power factor control), congestion management, and system restoration (islanded operation). By introducing new requirements based on new system needs, SOs encourage existing resources to offer/enhance their capabilities and signal new participants to invest in enhanced technology. These mandatory and non-remunerated requirements draw a baseline for the future design of ASs, which are sometimes quite complex, both technically and economically, to implement. 

The key difference between SIR and VIR is that the former is a natural and uncontrolled response, while the latter is an emulated response. Through adequate control strategies, VIR could be deliberately designed to deviate from the temporal shape of SIR, by responding based on the RoCoF, if the expected response proved to be more effective than just emulating SIR. PFR and FFR are both frequency-deviation-based reserves; however, the former relies on the slower governor response, whereas the latter comprises a faster response capability. In contrast to FFR, VIR is a RoCoF-based reserve~\cite{Poolla2019}. Notice that \textit{synthetic inertia} is generally used to describe VIR~\cite{Brisebois2011}. However, the term \textit{synthetic inertia} was coined under the deployment of GFL inverters, which have a natural response delay, and thus, FFR should be the preferred terminology. FRC differs from frequency regulation because the former reserves capacity to follow net load movements in future time dispatch intervals (minutes), while the latter reacts to continuous imbalances (seconds) in net load~\cite{Ela2017,Hu2018}. Since FRC is provided through the system dispatch, this AS differs from spinning reserve, which is reserved for support disturbances~\cite{Ahlstrom2015}. 

\subsection{Design of AS markets}

\begin{figure}[!h]
    \centering
     \tikzset{whtblock/.style={rectangle, rounded corners, draw=none, fill=teal!40, text width=23mm, text centered, minimum height=4em,font=\fontsize{8}{8}\selectfont}, 
     shdblock/.style={rectangle, rounded corners, draw=none, fill=lightgray!60, text width=23mm, text centered, minimum height=4em,font=\fontsize{8}{8}\selectfont},
     whtblock1/.style={rectangle, rounded corners, draw=none, fill=teal!40, text width=23mm, text centered, minimum height=1.8em,font=\fontsize{8}{8}\selectfont},
     shdblock1/.style={rectangle, rounded corners, draw=none, fill=lightgray!60, text width=23mm, text centered, minimum height=1.8em,font=\fontsize{8}{8}\selectfont},
    }
\begin{tikzpicture}[node distance = 0.5cm, auto]


\node [shdblock] (MP) {Mandatory\\[0.2ex] Provision};   
\node [whtblock, right=of MP, node distance=2.5cm] (SP) {Self-\\[0.2ex]Provision};
\node [whtblock, right=of SP, node distance=2.5cm] (BC) {Bilateral\\[0.2ex] Contracts};
\node [whtblock, right=of BC, node distance=2.5cm] (PT) {Public\\[0.2ex] Tendering};
\node [whtblock, right=of PT, node distance=2.5cm] (AU) {Auctions\\[0.5ex](monthly/weekly basis)};
\node [whtblock, right=of AU, node distance=2.5cm] (SM) {Day-Ahead/\\[0.2ex]Real-Time\\[0.3ex] Markets};


\node [shdblock, below=2 cm of MP.center] (NR) {Unpaid/\\[0.6ex]Regulated\,Price};
\node [shdblock1, below= 2 cm of SP.center] (RP) {Regulated\,Price};
\node [whtblock1, below= 0.5 cm of RP.center] (MB) {Market-Based};
\node [whtblock, below=2 cm of BC.center] (PAB) {Pay-As-Bid};
\node [shdblock1, below= 2 cm of PT.center] (RPT) {Regulated\,Price};
\node [whtblock1, below= 0.5 cm of RPT.center] (PABT) {Pay-As-Bid};
\node [whtblock, below=2 cm of AU.center] (PABA) {Pay-As-Bid};
\node [whtblock, below=2 cm of SM.center] (PAC) {Marginal\,Pricing/\\[0.4ex]Pay-As-Bid };


\draw [-{Stealth[length=2mm]}] (MP.south) |-($(MP.south)+(0,-0.4)$)-| (NR.north);
\draw [-{Stealth[length=2mm]}] (SP.south) |-($(SP.south)+(0,-0.4)$)-| (RP.north);
\draw [-{Stealth[length=2mm]}] (BC.south) |-($(BC.south)+(0,-0.4)$)-| (PAB.north);
\draw [-{Stealth[length=2mm]}] (PT.south) |-($(PT.south)+(0,-0.4)$)-| (RPT.north);
\draw [-{Stealth[length=2mm]}] (AU.south) |-($(AU.south)+(0,-0.4)$)-| (PABA.north);
\draw [-{Stealth[length=2mm]}] (SM.south) |-($(SM.south)+(0,-0.4)$)-| (PAC.north);

\end{tikzpicture}

    \caption{Procurement methods and pricing mechanisms. Rectangles shaded by gray color comprise non-market alternatives, while rectangles shaded by teal color identify market-based approaches.}
    \label{fig:Procurement_Pricing}
\end{figure}
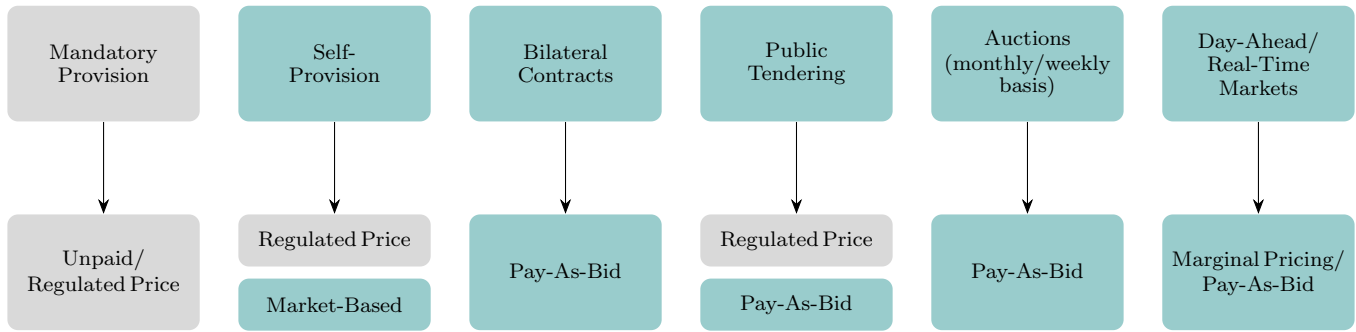

A few guiding practices should be followed for competitive procurement and efficient allocation of resources when designing AS markets, but no standard prescription conforms to the unique characteristics and historical practices of each power system. Every market defines its own set of ASs according to its infrastructural changes and operational requirements, leading to a continuous and complex redesign~\cite{Oren2001,Zhou2016}. A market redesign for a 100\% non-fossil future is most likely to be an improved market considering some fundamental existing rules, still valid under high shares of VRE, rather than a new market design built from scratch. An AS should follow a specific objective that reflects or anticipates future system needs. The AS product must be strictly defined, comprising technical and administrative requirements that eligible providers must follow. The procurement method, pricing mechanism, remuneration structure, and cost allocation scheme are key design variables that direct an AS market framework. Figure~\ref{fig:Procurement_Pricing} relates various procurement methods and pricing mechanisms.

ASs are acquired by SOs and market participants (utilities, generators, and consumers) with obligations in system security. Mandatory provision (remunerated or not) resembles a vertically integrated utility approach and aims to guarantee that certain capabilities must be provided. The market power of dominant agents is reduced, but additional costs to suppliers may cause unnecessary investments and the overproduction of resources. Self-provision allows market participants to use their portfolio to meet all, or a portion, of their AS obligations. A regulated price (RP) or market-based mechanism compensates the resources. To ensure a market framework, an AS must be competitively provided by a number of cost-efficient suppliers. Also, sufficient demand need for the AS must be ensured to justify the fixed operating costs~\cite{Pollitt2019}. Long-term bilateral contracts are helpful to hedge against the risk of insufficient reserve capacity or higher prices. Alternatively, monthly or weekly auctions procure reserves ahead of the spot market, paying suppliers according to the offers made. Public tendering processes are suitable for non-standard products, such as trials of emerging AS. 

Day-ahead market (DAM) and real-time market (RTM) comprise short-term platforms for AS acquisition that can be cleared based on pay-as-bid pricing (PBP) or marginal pricing (MP)~\cite{Glismann2017,ReboursEcon2007}. In the spot market, SOs can sequentially optimize energy and frequency-related AS (FR, SR, NSP, and RR)~\cite{Ma1999}. As energy and reserves are mutually exclusive, the foregone revenue by reserving capacity in AS markets, rather than selling electricity in the energy market, represents the lost opportunity cost (LOC) associated with reserve provision~\cite{Morey2001}. Scheduling and procuring energy in advance can result in poor allocation of the decisions and price signals~\cite{Divenyi2019}. Alternatively, co-optimizing energy and multiple reserves simultaneously optimizes the market products and, thus, adequately allocates and prices them, including the reimbursement of opportunity costs~\cite{Gonzalez2014}. Reserves follow a hierarchy in response time, where better-quality reserves (faster response) can replace low-quality (slower response) ones. Several SOs enforce this downward substitutable characteristic to provide appropriate price signals across the reserve categories. A price hierarchy ensures that prices decrease from higher to lower quality reserves, that is, from FR through contingency reserves (SR $>$ NSR $>$ RR)~\cite{Oren2001,Ela2016}. The higher LOC experienced by FR is due to frequent power output changes in short-time intervals, which reduce the revenue in the energy market and increase maintenance costs~\cite{Kirby2007}.

The remuneration structure reflects the costs incurred for AS provision. Since energy and capacity are different products, resources should submit separate offer prices for the creation of two distinct merit orders. Considering that energy and reserves are co-optimized, if a resource is selected to dispatch in real time, it should be paid for the electricity delivery. Also, if a resource is called to provide reserves, capacity reservation payments, which internalize the incurred opportunity costs, should be paid~\cite{Singh1999,Morey2001}. However, if energy and reserves are sequentially optimized, SOs could adopt availability payments to compensate units that reserve capacity within a predefined window for a later call. Furthermore, utilization payments reimburse units for the electricity delivered during reserve provision. Payment for performance is applied in specific ASs, such as frequency regulation, to encourage improved response. In general, the allocation of costs for the procured AS relies on a tariff, which is socialized across customers. However, a more economically efficient scheme should consider the cost causation principle, that is, those market participants who cause the costs to the system should pay those costs~\cite {Kirby2003ca}.   
A market-based framework is unsuitable for some ASs because of their specificities. The general practice of SOs shows that non-frequency-related ASs involve certain technical and economic barriers that must be overcome before adopting a competitive procurement mechanism~\cite{Anaya2018,Gracia2019}. Voltage control is highly sensitive to the grid location, which facilitates the abuse of some suppliers' market power. Moreover, a full AC power flow model must be performed to evaluate voltage support needs and price the service. However, such a model is non-linear and non-convex, which increases the computational complexity and makes its solution challenging for an RTM~\cite{Ela2016,Raineri2006}. Reactive power procurement is generally compulsory or agreed upon through bilateral contracts, and resources are compensated via cost-based payments or a provision tariff~\cite{FERC2005}. Black start resources are typically procured through long-term bilateral contracts~\cite{Isemonger2007,Kirschen2019}. The main hurdles in setting up a market for black start capability are the technical and locational restrictions of the providers. Not all generators have the desired technical capabilities to support power system restoration. If a generator is capable, it should be strategically located to restore the main feeders according to the restoration plan~\cite{Kirby1999}.    

\section{US Wholesale Electricity Market}

\begin{figure}[!h]
    \centering
    \includegraphics[]{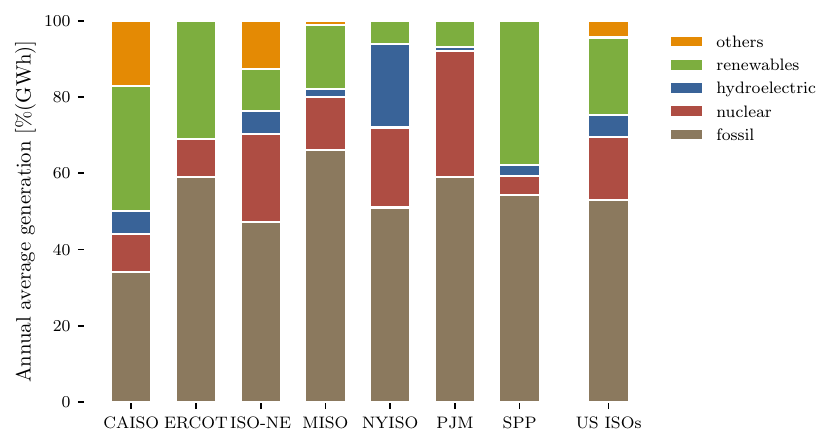}
    \caption{\justifying Annual average electricity generation [\%(GWh)] mix for US ISOs in 2022 (except CAISO: 2021). Renewables include: CAISO: solar, wind, geothermal, and biomass; ERCOT, ISO-NE, MISO, PJM, SPP: wind and solar; NYISO: wind, solar, biomass, and pumped hydropower.}
    \label{fig:EnergyMix}
    \vspace{-4mm}
\end{figure}

In the United States, SOs do not own transmission assets~\cite{Pollitt2008} and are called independent system operators (ISOs), while, historically, state-specific or, more broadly, regional transmission organizations (RTO), refer to greater footprints. ISOs and RTOs are very similar concepts. Currently, seven ISOs/RTOs oversee two-thirds of the US electricity load~\cite{FERCsite2019}: California ISO (CAISO), Electric Reliability Council of Texas (ERCOT), ISO New England (ISO-NE), Midcontinent ISO (MISO), New York ISO (NYISO), Pennsylvania-New Jersey-Maryland Interconnection (PJM), Southwest Power Pool (SPP). All ISOs operate under the jurisdiction of the Federal Energy Regulatory Commission (FERC) except ERCOT, due to historical reasons. Concepts underlying electricity markets have evolved from experiences on northeast regional power pools\footnote{Prior to restructuring, ISO-NE, NYISO, and PJM, were regional (tight) power pools, while CAISO and ERCOT were dominated by a few investor-owned utilities.}. However, utilities in the southeast, southwest, and northwest regions of the country remain vertically integrated. The electricity generation mix [\%(GWh)] of the seven US ISOs is shown in Fig.~\ref{fig:EnergyMix}.

All ISOs present a fossil fuel-based mix, mostly served by natural gas power plants, complemented mainly by nuclear, hydroelectric, and renewable energy power plants (wind and solar). Among the seven US ISOs, SPP has the most renewable mix, i.e., around 37.5\% of the total generation comes from wind power plants~\cite{ERCOTshare2023}. Similar to SPP, ERCOT presents high shares of wind generation, resulting in an annual average generation of 25\%. Additionally, solar power accounts for 6\% of the total supply~\cite{SPPshare2023}. Solar power is the dominant renewable in CAISO, reaching 16\% of the total generation, followed by 9\% for wind power generation. Imported electricity represents 17\% (others) of total generation in CAISO~\cite{CAISOshare2023}. MISO has an intermediate level of renewable generation (17\%), mostly supplied by wind power~\cite{MISOsite2022}. Renewables in ISO-NE represent 11\% of total generation while the contribution of net import electricity (others) is 14\%~\cite{ISONEshare2023}. NYISO (6.0\%) and PJM (7.0\%) present similar percent levels of renewable generation~\cite{NYISOshare2023}. The aggregate contribution of all ISOs results in an average of 20.4\% renewable generation. Each ISO coordinates system operation and clears a centralized DAM in its administrative region. The scheduling and dispatch process is shown in Fig.~\ref{fig:US_Design}.

\begin{figure}[!h]
    \centering
\tikzset{orgblock/.style={draw=none, rectangle, fill=my_orange, text centered, minimum height = 1.2cm},
base/.style={shape=signal, inner sep=0.2ex, align = center,
   decoration={shape backgrounds,shape size=0cm}},
   block/.style={base,signal from=west, signal to=east}
}

\begin{tikzpicture}[every node/.style, font=\fontsize{8.5}{8.5}\selectfont]
    
  \begin{scope}[node distance = 1mm,start chain=A going right,nodes={minimum height=2.5cm, on chain}]
  \node [block, fill= my_gray,signal from=nowhere, text width=2.1cm] (G) {Forward energy,\\[0.2ex] capacity,\,and\\[0.2ex] FTR markets};
  \node [block, fill= my_green, text width=2.1cm] (A) {Commitment/\\[0.2ex] Scheduling\,(SCUC) \\[0.5ex] Dispatch/\\[0.2ex] Pricing\,(SCED)};
  \node [block, fill= my_blue, align=center, text width=1.8cm] (D) {RUC\\[0.2ex] (update\\[0.2ex] commitment)};
  \node [block, fill= my_yellow, text width=1.8cm] {Look-Ahead\\[0.2ex] SCUC/SCED\\[0.5ex] Real-Time\\[0.2ex] SCUC/SCED};  
  \node [orgblock, above left = 6mm and 5mm of A.north, text width=1.6cm] (B) {Generator offers/\\parameters};
  \node [orgblock, above right = 6mm and -4mm of A.north, text width=1.1cm] (C) {Demand\\ bids};
  \node [orgblock,above left = 6mm and 8mm of D.north, text width=1.0cm] (E) {Load/\\VRE\\ forecast};
  \node [orgblock, right = 1.0mm of E.east, rectangle,text width=1.3cm] (F) {Generator\\ offers};

  \draw[-{Stealth[length=1.5mm]}] (B.south) -- ([xshift=-3.75em] A.north);
  \draw[-{Stealth[length=1.5mm]}] (C.south) -- ([xshift=0.74em] A.north);
  \draw[-{Stealth[length=1.5mm]}] (E.south) -- ([xshift=-3.74em] D.north);
  \draw[-{Stealth[length=1.5mm]}] (F.south) -- ([xshift=0.24em] D.north);
  \draw[-{Bar[scale=2]}] (-1.1,-2) -- (1.1,-2);
  \draw[-{Bar[scale=2]}] (1.1,-2) -- (7.9,-2);
  \draw[-{Stealth[length=3mm, width=2mm]}] (7.9,-2) -- (12.2,-2); 
  \node[minimum height=1mm] at (11, -2.5) {Time};
  \node[minimum height=1mm] at (9, -2.5) {D};
  \node[minimum height=1mm] at (8.2, -1.6) {Real-Time Market};
  \node[minimum height=1mm] at (3.0, -1.6) {Day-Ahead Market};
  \node[minimum height=1mm] at (-1.4, -1.6) {Forward\,Markets};
  \node[minimum height=1mm] at (4.5, -2.5) {D-1};
  \node[minimum height=1mm] at (-1.1, -2.5) {Years/Months};
  \end{scope}
\end{tikzpicture}
\vspace{-2mm}
       \caption*{Acronyms: FTR: financial transmission rights; SCUC: security-constrained unit commitment; SCED: security-constrained economic dispatch; RUC: reliability unit commitment.}
       \vspace{-2mm}
       \caption{Market design in the US relies on solving a complex scheduling and dispatch optimization problem, centralizing the balancing responsibility in the ISO.}
    \label{fig:US_Design}
    \vspace{-4mm}
\end{figure}

On a long-term basis, forward energy and capacity markets ensure resource adequacy, while financial transmission rights (FTR) allow market participants to hedge congestion charges associated with long-term contracts. In the DAM, producers submit detailed physical parameters and offers for energy and ASs. A three-part offer is the prevailing format in the energy market, which comprises incremental offers (MW-h, \$/MWh), a start-up fee (\$/start), and a no-load fee (\$/h)~\cite{Herrero2020}. The ISO gathers the demand bids and performs a centralized unit commitment, jointly optimizing energy and reserves\footnote{Differently from the other US ISOs, there is no day-ahead reserve product in ISO-NE. A forward reserve market run before day-ahead to provide reserve capacity for real-time physical delivery~\cite{ISONEfrm2019}.}, considering transmission network constraints and security requirements~\cite{FERC2014price}. The security-constrained unit commitment (SCUC) decides which resources should be available and what their output should be. Moreover, the security-constrained economic dispatch (SCED) determines the locational marginal prices (LMPs)~\cite{Ahlqvist2019,Lin2017}. Additionally, the ISO runs a reliability unit commitment (RUC) to ensure that sufficient energy and AS capacity are committed to serve the forecasted net load~\cite{Zhang2022}. After the DAM has closed, participants can inform the SO of any change in their operating plans, but cannot update their offers~\cite{Cramton2017}. In real time, a look-ahead SCUC or SCED (or both) assists the ISO dispatch decisions for evaluating future system conditions. In addition, a SCUC/SCED, typically 5-minute resolution, co-optimizes energy and reserves (except ERCOT\footnote{In ERCOT, the reserve levels obtained in DAM are held in RTM.}) to optimally allocate the resources in real time~\cite{FERC2014price}. The AS clearing prices are retrieved from the dual variables of each service constraint and reflect the incurred LOC. 

\subsection{AS Markets in ISOs/RTOs}

In the US, ISOs are the authority responsible for balancing the power system and acquiring ASs. Table~\ref{tab:US_ISOs} details the ASs procured in each ISO following the classification introduced in Fig.~\ref{fig:Taxonomy}, also highlighting the technical and market features of each AS. Historically, the ERCOT power system evolved synchronously independent of the west and east US interconnections. Reduced levels of inertia and weak interconnection motivated a \textit{synchronous inertial response} product proposal. Among the US ISOs, only ERCOT has identified a potential need for financial incentives to guarantee minimum inertia levels. However, creating a new service has not achieved a consensus among stakeholders~\cite{ERCOTinertia2014}. {Additionally, no ISO has specific guidance for GFM requirements or has defined an explicit VIR service. 

\begin{table}[!h]
    \tiny
    \centering
    \caption{US ISOs ancillary service markets structure reveals the emergence of a flexible ramping capability service and primary frequency response (PFR) requirement, emphasized by the red color.}
    \vspace{-2mm}
    \label{tab:US_ISOs}
    \begin{threeparttable}
    \begin{tabularx}{\linewidth}{@{\extracolsep{-12pt}}>{\hsize=1.35\hsize}X
                         *{7}{>{\hsize=0.95\hsize}X}} 
    \hline\\[-2.5ex]
    \textbf{}& \textbf{CAISO} & \textbf{ERCOT} & \textbf{ISO-NE} & \textbf{MISO} & \textbf{NYISO}  &
    \textbf{PJM} & \textbf{SPP}\\[0.5ex]\hline\\[-2.3ex]
    \textbf{Primary Frequency \newline Control} &  &  & & & & & \\[4ex]
    \textit{\textcolor{g1}{Product name}} & \textcolor{g1}{Primary\,Frequency \newline Response} & \textcolor{g1}{Responsive\newline Reserve\,Service}  & \textcolor{g1}{Primary\,Frequency \newline Response}  & \textcolor{g1}{Primary\,Frequency \newline Response} & \textcolor{g1}{Primary\,Frequency \newline Response} & \textcolor{g1}{Primary\,Frequency \newline Response} & \textcolor{g1}{Primary\,Frequency \newline Response}  \\[4ex]
    \textit{Procurement method} & Mandatory & PFR:Mandatory\newline Loads:DAM  & Mandatory & Mandatory & Mandatory & Mandatory & Mandatory  \\[4ex]
    \textit{Pricing mechanism} & Unpaid & PFR:Unpaid \newline Loads:MP  & Unpaid  & Unpaid & Unpaid & Unpaid & Unpaid  \\[2ex]
    \textbf{Secondary Frequency \newline Control (continuous)} & & & & & & &\\[4ex]
    \textit{Product name} & Regulation up/down & Regulation\,Service\newline \textcolor{g1}{Fast\,Responding} & Regulation &    Regulation& Regulation &    Regulation &    Regulation \\[4ex]
    \textit{Separate\,up/down\,products} & \checkmark & \checkmark & \xmark &   \xmark &    \xmark &    \xmark &\checkmark\\[2ex]
    \textit{Procurement method} &   DAM/RTM & DAM & RTM &   DAM/RTM  & DAM/RTM & RTM &  DAM/RTM\\[2ex]
    \textit{Pricing mechanism} & capacity:\,MP \newline performance: mileage\,offer\,of marginal provider & capacity:\,MP \newline performance:\,non-remunerated & capacity:\,MP \newline performance:\newline mileage\,offer\,of marginal provider &   capacity:\,MP \newline performance: mileage\,offer\,of marginal provider & capacity:\,MP \newline performance: mileage\,offer\,of marginal provider & capacity:\,MP \newline performance: mileage\,offer\,of marginal provider & capacity:\,MP \newline performance: mileage\,offer\,of marginal provider\\[10ex] 
    \textbf{Secondary Frequency \newline Control(event-driven)} & & & & & & & \\[4ex] 
    \textit{Product name} & Spinning/Non-Spinning Reserve & Contingency\newline Reserve & 10-Min\,Synchro-nized/10-Min Non-Synchronized & Spinning/Supple-mental Reserve &  10-Min Spinning/10-Min Non-Synchronized &   Synchronized/\newline Non-Synchronized Reserve &    Spinning/Supple-mental Reserve\\[7ex]
    \textit{Procurement method} & DAM/RTM & DAM &   Forward auction/\newline RTM    & DAM/RTM & DAM/RTM & RTM & DAM/RTM\\[4ex]
    \textit{Pricing mechanism} & MP &  MP & MP & MP &   MP  & MP\,(non-sync.: cost-based) & MP\\[2ex]
    \textbf{Tertiary Frequency \newline Control (event-driven)} & & & & & &\\[4ex] 
    \textit{Product name} & - & Non-Spinning\newline Reserves\,(30-Min) &   30-Min\,Operating Reserves &    - & 30-Min\,Spinning/ 30-Min\,Non-Synchronized &    Secondary\,Reserve &    -\\[7ex]
    \textit{Procurement method} &   -   & - &   Forward auction/\newline RTM &  - & DAM/RTM &   - & -\\[4ex]
    \textit{Pricing mechanism} & - & - & MP & - & MP &  - & -\\[2ex]
    \textbf{Economic Dispatch} & & & & & & &\\[4ex]
    \textit{\textcolor{g1}{Product name}} & \textcolor{g1}{Flexible Ramping} \newline \textcolor{g1}{Product} & - & - & \textcolor{g1}{Ramp Capability} \newline \textcolor{g1}{Product} & - & - & \textcolor{g1}{Ramp Capability} \newline \textcolor{g1}{Product}\\[4ex]
    \textit{Separate\,up/down\,products}    & \checkmark &  -   & - & \checkmark & - & - & \checkmark\\[2ex]
    \textit{Procurement method} &   FMM/RTM &   - & - & DAM/RTM & - & - & DAM/RTM\\[2ex]
    \textit{Pricing mechanism}  & MP &  - & - & MP &    - & - & MP\\[2ex]
    \textbf{Voltage Control} & & & & & & &\\[2ex]
    \textit{Product name} & Voltage Support & Voltage Support\newline Service   & Reactive Supply\newline Service & Reactive Supply \newline Service    & Voltage Support \newline Service &    Reactive Service &  Reactive Supply \newline Service\\[4ex]
    \textit{Procurement method} &   Mandatory & Mandatory & Mandatory & Mandatory & Mandatory & Mandatory & Mandatory\\[2ex]
    \textit{Remuneration structure} & Utilization \newline payment  & Utilization\newline payment & Capability/Utili-zation payments & Capability\newline payment & Capability/Utili-zation payments &  Capability/Utili-zation payments & Utilization\newline Payment \\[2ex]
    \textbf{System Restoration} & & & & & & & \\[2ex]
    \textit{Product name} & Black Start\newline Capability & Black Start\newline Service    & Black Start\newline Service & Black Start\newline Service &   Black Start\newline Capability & Black Start\newline Service &  -\\[4ex]
    \textit{Procurement method} & Service agreement \newline (resource/ISO) & Competitive \newline bidding process  & Service agreement \newline (resource/ISO) & Service agreement \newline (resource/ISO) & Service agreement \newline (resource/ISO) & Service agreement \newline (resource/ISO) & Not procured\\[5ex]
    \textit{Remuneration structure} & Service payment \newline (recovey costs) &    Hourly ``standby \newline fee" payment &    Standard or \newline station-specific rate payment &  Fixed + Variable \newline
 + Training + \newline Compliance costs &   Service payment \newline
 (recovey costs)    & 1.1 $\cdot$  (Fixed \newline + Variable + Fuel \newline
 + Training costs) &    -\\[7ex]
 \textit{Cost Recovery} & OATT: consumers & OATT: consumers &   OATT: consumers &   OATT: consumers &   OATT: consumers &   OATT: consumers &   OATT: consumers\\[0.5ex]
     \hline
\end{tabularx}
\end{threeparttable}
\begin{tablenotes}
    \scriptsize
    \item Sources:~\cite{Ela2016},~\cite{Ela2019},~\cite{Zhou2016},~\cite{CAISOtariff2020},~\cite{ERCOTtariff2021},~\cite{ISONEtariff2020},~\cite{MISOtariff2023},~\cite{NYISOtariff2023},~\cite{PJMtariff2010},~\cite{SPPtariff2012}. FMM: fifteen-minute market, OATT: open access transmission tariff.
\end{tablenotes}
\end{table}

The primary frequency response has been considered as a byproduct of SG operation in all ISOs for decades. However, after the NERC BAL-003-1.1 standard~\cite{NERCBAL32019} and FERC Order 842~\cite{FERC842}, primary frequency response evolved into an obligatory capability. The standard mandates that balancing authorities demonstrate sufficient primary frequency response capability. The order imposes primary frequency response capability requirements for all newly interconnecting large and small units, synchronous (except nuclear power plants) or non-synchronous (including ESS). In ERCOT, \textit{responsive reserves} are being implemented to aggregate slower and faster reserves related to the primary frequency response, complying with the NERC standard BAL-001-TRE-1~\cite{NERCBALercot}. Besides the autonomous governor response from SG (slower reserves), ERCOT procures \textit{fast frequency response} capability from interruptible loads with under-frequency relays, and is introducing the procurement from BESS~\cite{Du2020,Matevosyan2017}. The latter should respond automatically within 0.25\,s at 59.85\,Hz threshold while maintaining a full response for at least 15 minutes~\cite{Matevosyan2019,Li2018,ERCOTffr2020}. Only ERCOT in the US is introducing a payment for SG governor response. The ISO intends to compensate for the availability of delivery reserve capacity (\$/MW) through a market framework~\cite{ERCOTManagement2023,ERCOTPRF2018}.

Frequency regulation is designed with separate products for upward (regulation up) and downward movements (regulation down) in CAISO, ERCOT, and SPP. In PJM, two products have been created to separate traditional units with slow ramp rates (controlled by the RegA signal) from new fast ramp rate resources, such as BESS (controlled by the RegD signal)~\cite{PJMManualAncillary2022,PJMRegulation2017}. Notice that frequency-based ASs are procured in the DAM in ERCOT and are physically binding in real time. If necessary, ERCOT may procure additional reserves in real time through the \textit{supplemental ancillary services market}~\cite{ERCOTAhead2023}. Apart from compensation for available capacity through marginal pricing, frequency regulation providers are also remunerated based on their ability to follow the AGC signal, complying with FERC Order 755~\cite{FERCO}. The exception is ERCOT, which does not monitor the accuracy of frequency regulation providers~\cite{Byrne2015}. Prices for performance are set as the mileage offer of the marginal capacity provider~\cite{Chen2015}. \textit{Fast-responding regulation} is a sub-product of the \textit{regulation service} in ERCOT. This new AS was launched under a pilot project and is a tailor-made AS to reward the benefits of the fast-ramping capability of ESS. Resources must provide regulation capacity within one second after the ERCOT signal or after independent identification of a trigger frequency~\cite{Peydayesh2018,ERCOTManagement2023}. 

All ISOs procure synchronized resources that are fully available within 10 minutes to supply spinning reserves. Additionally, non-synchronized resources capable of responding within 10 minutes are eligible to provide non-spinning reserves. ISOs remunerate the availability of capacity based on marginal prices, except for the \textit{non-synchronized reserve} in PJM, which is cost-based. Some ISOs acquire an additional 30-minute spinning and non-spinning reserve to serve as replacement reserves. In NYISO, the sum of the \textit{total 10-minute reserve} and the \textit{total 30-minute reserve} must be greater than or equal to twice the largest single contingency~\cite{NYISOManualAncillary2022,NYISOReserve}. In PJM, the \textit{day-ahead scheduling reserve market} procures 30-minute reserves (so-called \textit{secondary reserves}). PJM determines the requirements for this product based on load forecasting for the following operational day, but it does not impose performance obligations in real time~\cite{PJMReserve2016,Monitoring2022}. Additionally, to incentivize the response of flexible resources during a reserve shortage, ISOs in US are gradually introducing the operating reserve demand curve (ORDC) approach, which ensures enhanced price signals when available capacity is scarce. 

Flexible ramping capability is procured in CAISO, MISO, and SPP when the net load ramping
requirements exceed the ability of dispatched units to follow net load, thus, incentivizing flexible resources to reserve capacity~\cite{Ela2012}. In CAISO, high shares of solar power cause an upward ramp in the morning and a downward ramp in the evening (so-called duck curve). The \textit{flexible ramping product} in CAISO emerges after including a constraint in the real-time dispatch~\cite{CAISO2011}. Currently, CAISO procures upward and downward capacity in the \textit{fifteen-minute market} (FMM) and RTM markets to provide the ramping capability for the next 15-minute interval, consisting of three consecutive 5-minute intervals. To determine the procurement and shadow prices, CAISO uses a demand curve based on the net demand forecast uncertainty for the next time interval, simultaneously extracting the VRE and demand forecast errors~\cite{CAISOreport2019}. In MISO, wind power is the dominant renewable, and the \textit{ramp capability product} is procured over 10 minutes in the day-ahead and real-time markets~\cite{MISOramp2023}. Similar to MISO, SPP has observed an accelerated growth of wind generation. Thus, a \textit{ramp capability product} was launched in 2022 considering a 20-minute interval~\cite{SPPramp2022}.

A voltage control-related AS is mandatory for synchronous, and all newly interconnecting non-synchronous (wind and solar), units, according to FERC Order 827~\cite{FERC827}. Additionally, large and small facilities connected to the transmission system should provide dynamic reactive power support, complying with the voltage ride-through capability requirement stated in FERC Order 828~\cite{FERC828}. Cost-based utilization, or capability (or both) payments, reimburse the costs of providers~\cite{FERCvs2014}. Black-start resources are procured through service agreements, being compensated for cost-based rates to recover incurred costs. Currently, there is no black start capability procurement in SPP. ERCOT applies a competitive biannual auction to define resources with the lowest costs~\cite{Sun2011}. To reimburse AS costs, a transmission tariff (open access transmission tariff; OATT) is applied to transmission customers in all US ISOs~\cite{FERCoatt2013}.

\section{European Wholesale Electricity Markets}

In Europe, SOs are known as transmission system operators (TSOs), and unlike the US, are allowed to own transmission assets. European countries typically adopt a decentralized DAM with sequential optimization of energy and reserves\footnote{Italy and Spain co-optimize energy and reserves in DAM.} and zonal prices. These energy markets mainly rely on self-dispatch, where suppliers should communicate their operating plan to TSOs, and they can also decide the production of each unit. By sharing a greater responsibility in planning system operation with providers, TSOs tend to be less active than the ISOs of centralized DAMs, which implies a market based on financial exchanges~\cite{Ahlqvist2022}. Figure~\ref{fig:EU_Design} shows the organization of European wholesale electricity markets.  

\begin{figure}[!h]
    \centering

\tikzset{orgblock/.style={draw=none, rectangle, fill=my_orange, text centered, minimum height = 1.5cm},
base/.style={shape=signal, inner sep=0.2ex, align = center, decoration={shape backgrounds,shape size=0cm}},
block/.style={base,signal from=west, signal to=east},
}

\begin{tikzpicture}[every node/.style, font=\fontsize{8}{8}\selectfont]

  \begin{scope}[node distance = 1mm,start chain=A going right,nodes={minimum height=2.5cm, on chain}]
    \node [block, fill= my_gray,signal from=nowhere,text width=2.1cm] (G) {Forward energy,\\[0.2ex] FTR\,markets \\ [0.2ex]capacity\,markets/\\[0.2ex]mechanisms};
    \node [block, fill= my_green,text width=3.2cm] (A) {Market coupling\\(EUPHEMIA)/\\[0.2ex] TSO redispatch};
    \node [block, fill= my_blue, text width=2cm] (D) {Continuous\\[0.2ex] trading\,and \\[0.2ex] TSO\,redispatch};
    \node [block, fill= my_red, text width=2.4cm] (E) {Activation of\\[0.2ex] balancing energy/\\[0.2ex] TSO\,balancing};
    \node [block, fill= my_yellow, text width=1.3cm] {Settles\,BRP \\[0.2ex]imbalances}; 
    \node [signal, align=center, fill= my_purple, text width=6.3cm,minimum height=0.7cm, below left=1mm  and 2.5mm of A.south east] (Z) {Forward reservation capacity in\\ balancing (capacity) market};
    \node [orgblock, above right = 6mm and -4mm of A.north west, rectangle,text width=1.3cm] (B) {Generator\\[0.2ex] offers};
    \node [orgblock, right = 0.5mm of B.east, rectangle,text width=1.1cm] (C) {Demand\\[0.2ex] bids};
    \node [orgblock, above right = 0mm and 0.5mm of C.south east, rectangle,text width=1.8cm, align=center] (H) {Cross border \\[0.1ex] capacity\\[0.2ex] allocation \\[0.2ex] (TSOs)};
  \node [orgblock, above left = 6mm and 5mm of D.north, rectangle,text width=1.1cm] (E) {VRE\\[0.2ex] forecast};
  \node [orgblock, right = 0.5mm of E.east, rectangle,text width=1.1cm] (F) {Updated\\[0.2ex] offers};
  \draw[-{Stealth[length=2mm]}] (B.south) -- ([xshift=-6.54em] A.north);
  \draw[-{Stealth[length=2mm]}] (C.south) -- ([xshift=-2.59em] A.north);
  \draw[-{Stealth[length=2mm]}] (H.south) -- ([xshift=2.02em] A.north);
  \draw[-{Stealth[length=2mm]}] (E.south) -- ([xshift=-3.10em] D.north);
  \draw[-{Stealth[length=2mm]}] (F.south) -- ([xshift=0.63em] D.north);
  \draw[-{Bar[scale=2]}] (-1.1,-2.9) -- (1.1,-2.9);
  \draw[-{Bar[scale=2]}] (1.1,-2.9) -- (5.7,-2.9);
  \draw[-{Bar[scale=2]}] (5.7,-2.9) -- (9.1,-2.9);
  \draw[-{Bar[scale=2]}] (9.1,-2.9) -- (13,-2.9);
  \draw[-{Stealth[length=3mm, width=2mm]}] (13,-2.9) -- (16.8,-2.9); 
  \node[minimum height=1mm] at (15.4, -3.3) {Time};
  \node[minimum height=1mm] at (13, -2.5) {Imbalance Settlement};
  \node[minimum height=1mm] at (-1.4, -2.5) {Forward\,Markets};
  \node[minimum height=1mm] at (1.4, -2.5) {Day-Ahead Energy Market};
  \node[minimum height=1mm] at (9, -2.5) {Balancing (Energy) Market};
  \node[minimum height=1mm] at (5.6, -2.5) {Intraday Energy Market};
  \node[minimum height=1mm] at (13.8, -3.3) {D+1};
  \node[minimum height=1mm] at (6.8, -3.3) {D};
  \node[minimum height=1mm] at (10.2, -3.3) {D};
  \node[minimum height=1mm] at (3.0, -3.3) {D-1};
  \node[minimum height=1mm] at (-1.2, -3.3) {Years/Months};
  \end{scope}
  
\end{tikzpicture}
\vspace{-2mm}
    \caption*{Note: FTR: financial transmission rights; BSPs: balancing service providers; BRPs: balancing responsible parties.}
    \vspace{-2mm}
       \caption{European market design typically focuses on self-scheduling of units and decentralized balancing responsibility.}
    \label{fig:EU_Design}
    \vspace{-4mm}
\end{figure}

Capacity mechanisms (strategic reserve or capacity payment) have been introduced in Europe to facilitate the integration of VRE~\cite{Kozlova2022}. In the day-ahead energy market, participants typically submit a simple price-quantity offer to a financial platform known as power exchange. Considering the available cross-border capacity, as informed by the TSOs, the EUPHEMIA algorithm clears the single day-ahead energy market by coupling several wholesale markets across Europe. The results are the electricity price and the net position for each bidding zone~\cite{Euphemia2020}. TSOs should reserve capacity beforehand through auctions covering different time frames (yearly, monthly, weekly, daily) in their national \textit{balancing capacity market} from \textit{balancing service providers} (BSPs). TSOs have the final responsibility to balance their control area in real time~\cite{Veen2016,ENTSOE2018}. Market participants are also responsible for balancing the system, and are incentivized to self-balance. To do so, a market participant must be connected to a \textit{balancing responsible party} (BRP), which is financially responsible for maintaining balanced portfolios. BRPs send the generation and load schedules to the TSOs for planning the operational day~\cite{Poplavskaya2020,Roumkos2022}. TSOs run a sequential optimization of energy and reserves to verify the feasibility of their dispatch~\cite{Batlle2013}. If security constraints are violated, a redispatch is carried out~\cite{Ahlqvist2019,Matenli2016}. Using updated information from the VRE forecast on the operating day, market participants can update their positions in the intraday market (IDM) by continually buying and selling energy (continuous trading). In real time, TSOs must balance unforeseen disturbances by activating balancing energy quantities previously secured in the balancing capacity market. BSPs can also offer their energy availability in the \textit{balancing energy market} for real-time operation. The imbalance settlement determines the imbalance charge that BRPs must pay according to their deviation from the schedule~\cite{Cauret2019,ENTSOE2018}.

The annual average electricity [\%(GWh)] generation mix of several European countries is shown in Fig~\ref{fig:EnergyMixEurope}~\cite{Eurostat2023}. Renewables represent 18.2\% of European generation (selected countries) and are complemented by conventional fossil fuels (mainly oil and gas) and hydroelectric generation. Countries such as Norway and Austria possess high hydroelectric generation, while France and Belgium have a significant share of nuclear energy. By contrast, Denmark, Germany, and Ireland have higher renewable generation.    

\begin{figure}[!h]
    \centering
    \includegraphics[width=1.0\columnwidth]{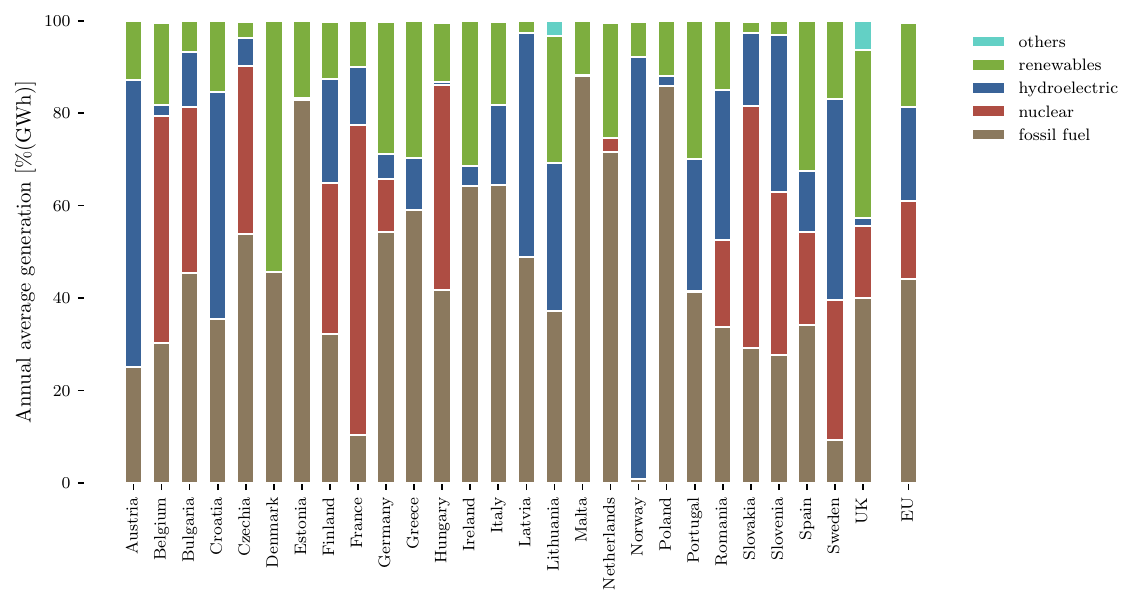}
        \vspace{-8mm}
       \caption{Annual average electricity [\%(GWh)] generation mix of different European countries (2021). Renewables include solar, wind, geothermal, tidal, wave, and ocean sources.}
    \label{fig:EnergyMixEurope}
\end{figure}

\subsection{Toward an Internal Electricity Market}

Since the 1990s, the European Union has considered building a single wholesale electricity market. To achieve this goal, the European Commission created an agreement in 1996 to set standard rules for the internal electricity market~\cite{Karan2011}. Nevertheless, insufficient transmission capacity resulted in the creation of regional electricity markets. In these markets, TSOs have procured reserves at the national level, potentially overestimating the necessary amount and increasing costs. Integrating several European markets requires product standardization and specific trading platforms. The European Network of Transmission System Operators for Electricity (ENTSO-E) sets a uniform framework of balancing services, that is, primary, secondary, and tertiary control-related services, being aligned with balancing exchange platforms~\cite{Meeus2020}, as follows:

\begin{itemize}[(1)]
    \item \textit{Frequency Containment Reserve (FCR)}: This service is constantly activated (outside the deadband between 49.99 Hz to 50.01 Hz) to contain frequency deviations. A local and automatic response, typically up to 30\,seconds, which should be sustained for 15 minutes, is requested to stabilize the frequency fluctuation~\cite{ENTSOE2018}. 
\end{itemize}

\begin{itemize}[(2)]
    \item \textit{Frequency Restoration Reserve (FRR)}: Two reserves comprise the FRR product: \textit{(i)} automatically activated FRR (aFRR) is a continuous reserve, while \textit{(ii)}  semi-automatic or manual FRR (mFRR) is a discrete reserve. These reserves should be available from 30\,seconds to 15 minutes, and conventionally maintained for hours. aFRR aims to replace FCR and restore the frequency to its nominal value~\cite{ENTSOE2018,Cauret2019}. A European-wide sizing and procurement of aFRR and mFRR are facilitated by the balancing platforms PICASSO ({P}latform for the {I}nternational {C}oordination of {A}utomated {F}requency {R}estoration and {S}table {S}ystem {O}peration) and MARI ({M}anually {A}ctivated {R}eserves {I}nitiative)~\cite{ENTSOEpic2022,ENTSOEmari2022}.
\end{itemize}

\begin{itemize}[(3)]
    \item \textit{Replacement Reserve (RR)}: This service is semi-automatic or manually activated within 15 minutes or more. The TERRE ({T}rans {E}uropean {R}eplacement {R}eserves {E}xchange) platform integrates the European RR markets. Notice that the procurement of RR is optional regarding the target model defined by ENTSOE~\cite{ENTSOEterre2022}.    
\end{itemize}

\begin{itemize}[(4)]
    \item \textit{Imbalance Netting (IN)}: To avoid simultaneous activation of aFRR in the opposite direction, TSOs are also responsible for maintaining an area control error close to zero, and correcting the input of aFRR accordingly. To this end, ENTSO-E implemented a platform called IGCC (International Grid Control Cooperation) for TSOs to exchange their real-time imbalances~\cite{Contu2019,ENTSOEimbal2022}.    
\end{itemize}

\subsection{AS Markets in Europe}

AS markets in Europe show many similarities but also some important differences. We analyze the AS markets in different European regional electricity markets, selecting power systems with high shares of renewables. The main findings are summarized for comparison in Table~\ref{tab:EU_countries}.

\begin{table}[!h]
    \fontsize{5.9}{8}\selectfont
    \centering
    \caption{In Europe, AS markets are rapidly evolving in GB, Ireland, and in the Nordic power system. The red color emphasizes recent innovations, such as the inclusion of a fast frequency response product.}
    \vspace{-2mm}
    \label{tab:EU_countries}
    \begin{threeparttable}
    \begin{tabularx}{\linewidth}{@{\extracolsep{-12pt}}>{\hsize=1.2\hsize}X >{\hsize=0.8\hsize}X >{\hsize=0.9\hsize}X >{\hsize=0.9\hsize}X >{\hsize=0.8\hsize}X >{\hsize=0.6\hsize}X} 
    \hline\\[-2.5ex]
    \textbf{}& \textbf{Great Britain} & \textbf{Ireland\,(all island)} & \textbf{Germany} & \textbf{Denmark} & \textbf{Sweden}\\[0.5ex]\hline\\[-2.3ex]
    \textbf{Inertial\,Response} & & & & &  \\[2ex]
    \textit{\textcolor{g1}{Product name}} & - & \textcolor{g1}{Synchronous\,Inertial\newline Response} & - & - & -\\[4ex]
    \textit{Procurement method} & - & Tender & - & - & - \\[1ex]
    \textit{Pricing mechanism} & - & RP & - & - & - \\[1ex]
    \textbf{Primary Frequency Control} & & & & &  \\[2ex]
    \textit{\textcolor{g1}{Product name}} & \textcolor{g1}{Dynamic\,Containment/\newline Moderation/Static\newline Recovery} & Primary\,Operating \newline Reserve/\textcolor{g1}{FFR}  & Primary Control \newline Reserve  & FCR-N,\,\textcolor{g1}{FCR-D\newline/FFR} & FCR-N,\,\textcolor{g1}{FCR-D\newline/FFR}  \\[7ex]
    \textit{Procurement method} & Market-based & Tender  & Market-based  & Market-based & Market-based\\[2ex]
    \textit{Pricing mechanism} & MP & RP  & MP  & MP & PBP  \\[1ex]
   \textbf{Secondary Frequency \newline Control (continuous)} & & & & & \\[4ex]
   \textit{Product name} &  \textcolor{g1}{Dynamic Regulation} & Secondary\,Operating\newline Reserve & Secondary Control \newline Reserve  & aFRR & aFRR  \\[4ex]
   \textit{Separate\,up/down\,products} & \checkmark &  \xmark & \xmark &   \checkmark & \xmark \\[2ex]
   \textit{Procurement method} & Market-based & Tender & Market-based & Market-based & Market-based  \\[2ex]
   \textit{Pricing mechanism} & MP & RP & PBP & PBP & PBP \\[2ex] 
   \textbf{Secondary Frequency \newline Control(event-driven)} & & & & & \\[4ex] 
   \textit{Product name} &  \textcolor{g1}{Quick/Slow Reserve} & Tertiaty\,Operating\newline Reserve 1/2 & Minute reserve & mFRR &  mFRR \\[4ex]
   \textit{Procurement method}  & Market-based &    Tender & Market-based & Market-based &  Market-based\\[2ex]
   \textit{Pricing mechanism} & MP &  RP &  PBP & MP &  PBP \\[2ex]
   \textbf{Tertiary Frequency \newline Control (event-driven)} & & & & \\[4ex] 
   \textit{Product name} &  - & Replacement\,Reserves \newline Synchronized/\newline Desynchronized & - & - & - \\[8ex]
   \textit{Procurement method} & - & Tender & -  & - &  - \\[2ex]
   \textit{Pricing mechanism} & - & RP & - & - & - \\[2ex]
   \textbf{Economic Dispatch} & & & & & \\[4ex]
   \textit{\textcolor{g1}{Product name}} &  - &\textcolor{g1}{Ramping Margin 1/3/8} & - & - & - \\[2ex]
   \textit{Procurement method} & - & Tender &   - & - & - \\[2ex]
   \textit{Pricing mechanism}   & - & RP &  - & - & - \\[2ex]
   \textbf{Voltage Control} & & & & & \\[2ex]
   \textit{Product name} &  Obligatory/Enhanced\newline Reactive\,Power\,Service & Steady-State\,Reactive Power,\,\textcolor{g1}{Dynamic\,Reactive Response,\,Fast\,Post-Fault Active Power Recovery}   & Static\,Voltage\,Stability \newline Dynamic\,Reactive\,Power  & Ordering Reactive\newline Reserve & Reactive Power \\[10ex]
    \textit{Procurement method} & Obligatory:\,Mandatory\newline Enhanced:\,Tender &    Mandatory & Mandatory or \newline Bilateral Contracts & Mandatory or \newline Bilateral Contracts & Mandatory or \newline Bilateral\,Contracts \\[4ex]
    \textit{Pricing Mechnism} & RP & RP & Contracts:RP & Contracts:RP & Contracts:RP  \\[2ex]
    \textbf{System Restoration} & & & & &  \\[2ex]
    \textit{Product name} & Black Start & Black Start & Black Start &   Black Start & Black Start \\[2ex]
    \textit{Procurement method} &  Tender   & EirGrid:Tender or \newline Bilateral Contract\newline SONI:Mandatory & 50\,Hertz:Bilateral &  Tender or\newline Bilateral Contract & Mandatory    \\[8ex]
    \textit{Remuneration structure} & Availability,\,testing,\newline warming,\,investment,\newline feasibility study costs  &  EirGrid:Availability,\newline testing,\,investment\,costs\newline SONI:RP & Fixed annual payment & Availability,\,testing,\newline investment\,activation  & Unpaid\\[8ex]
 \textit{Cost Recovery (capacity)} & FCR:BRP\newline mFRR:BRP\newline  &    FCR:consumers\newline aFRR:consumers\newline mFRR:consumers\newline RR:consumers &  FCR:consumers\newline aFRR:consumers\newline mFRR:consumers\newline &FCR:consumers\newline aFRR:consumers\newline mFRR:consumers\newline & FCR:consumers/BRP\newline aFRR:consumers/BRP\newline mFRR:consumers/BRP\newline \\[0.5ex]
     \hline
\end{tabularx}
\end{threeparttable}
\begin{tablenotes}
\scriptsize
\item Sources:~\cite{ENTSOEsurvey2021},~\cite{NGroad2019},~\cite{NGreserve2021},~\cite{EirGridproposal2013},~\cite{EirGrid2020},~\cite{Energiforsk2020},~\cite{Nordic2019},~\cite{Nordic2022},~\cite{Elia2018}. RP: regulated price; MP: marginal pricing; PBP: pay-as-bid pricing.
\end{tablenotes}
\end{table}

\subsubsection{Great Britain}

Great Britain's (England, Scotland, and Wales) power system is an island power system interconnected by HVDC links to other countries, which enhances the negative effects of low inertia levels. Great Britain's TSO, National Grid ESO, was the first SO to determine grid code requirements for GFM inverters~\cite{NGgfm2022}, which opens a path for VIR and system strength service provision. The AS markets are evolving from an intricate framework, with an overlap of services, toward a simpler and rational design, phasing out some frequency-related services~\cite{NGfut2020}. National Grid ESO introduced three new ASs, comprising a new suite of services designed for continuously tracking system frequency variations and replacing the existing \textit{dynamic firm frequency response} and \textit{enhanced frequency response} services. \textit{Dynamic containment} is a fast-acting (1 second) post-fault response to manage higher RoCoF after a disturbance, associated with SGs being displaced by VRE generation. In contrast, \textit{dynamic regulation} and \textit{dynamic moderation} are pre-fault services, that is, they aim to correct the system frequency before it moves outside the operational limit specified for the service. The former is a continuous response to stabilize small and continuous deviations in the operational frequency range~\cite{NGroad2019}. The latter is an additional fast response to manage larger imbalances and arrest the system frequency, responding within 1 second~\cite{NGroad2019}. Both \textit{dynamic containment/moderation} services are well-suited for BESS, whereas \textit{dynamic regulation} accommodates the capabilities of traditional suppliers~\cite{NGroadmap2022}. The \textit{static firm frequency response} service is provisionally renamed as \textit{static response} and comprises load shedding when a target frequency setpoint is reached~\cite{NGffr2023}. To supply static and dynamic services, providers can submit offers in a single clearing price day-ahead auction. 

A \textit{quick reserve} service has been designed as a pre-fault, bi-directional, and manually activated service within one minute after TSO instruction, to follow frequency deviations during normal conditions. \textit{Slow reserve} has been designed as a post-fault, bi-directional, and manually activated service {within 15 minutes after TSO instruction}, to restore the frequency after large imbalances~\cite{NGreserve2022,NGslow2022}. Quick and slow reserves intend to replace the \textit{short-term operating reserve} and \textit{fast reserve} services in the coming years~\cite{NGreserve2021}.

Integration into the TERRE platform was envisioned before Brexit, but National Grid ESO officially leaves the project in December 2022, leading to uncertainties about the future of a replacement reserve service in Great Britain~\cite{ENTSOE2023,NGroadmap2022}.  
Voltage control is a mandatory service compensated by a utilization payment~\cite{NGors2023}. Providers capable of supplying an additional reactive power response can offer this capability through a tender arrangement~\cite{NGers2023}. In 2020, National Grid ESO introduced a competitive pay-as-bid process for procuring black start resources. Availability and other minor payments compensate providers~\cite{NGblack2021}. The operating costs of the ASs are recovered from a system charge applied to BRPs.

\subsubsection{Ireland/Northern Ireland}

Similar to Great Britain, Ireland is an island power system with non-synchronous interconnections to its neighbors, and newly designed ASs facilitate the integration of high shares of wind (and solar) power. EirGrid and SONI are the first SOs to introduce a service related to the provision of synchronous inertia capability. \textit{Synchronous inertial response} aims to incentivize SGs to reduce their stable minimum power output, enabling the dispatch of other units to ensure a minimum inertia level~\cite{Tuohy2019}. The service is procured on a regular tender process, and providers receive a payment that considers the available rotational energy and minimum generating level of the unit~\cite{EirGridproposal2013}. \textit{Fast frequency response} is procured from resources capable of responding within two seconds to contain the frequency decay, but with financial incentives to provide a faster response.

\textit{Primary, secondary, and tertiary operating reserves} and \textit{replacement reserves (synchronized and desynchronized)} are the existing services for frequency control~\cite{Delaney2020}. Three ramping services schedule available ramping capability over 1, 3, and 8 hours to manage uncertainties associated with VRE forecasts. Unlike similarly named services introduced in CAISO and MISO, the \textit{ramping margin} created by EirGrid and SONI relies on a long-term schedule and focuses only on upward movement, given that generators can be conveniently requested to switch offline or VRE can be curtailed if the available VRE generation greatly exceeds the forecast~\cite{EirGridramp2021}. 

\textit{Steady-state reactive power} is the current AS for voltage control under normal conditions. Since significant shares of VRE connected to the distribution system are displacing SG units, a reduction in reactive power capability is noticed. Particularly, geographic locations, far from consumer centers, and thus, with low demand and weaker networks, have experienced an increase in magnitude and frequency of occurrence of low voltage deviations in transmission buses~\cite{Nolan2020}. In addition, the scarcity of dynamic reactive power capability, associated with the reduction of synchronizing torque due to fewer online synchronous units, is anticipated at very high (+70\%) instantaneous non-synchronous shares. \textit{Dynamic reactive response} is designed as a new AS to increase the transient reactive power response and mitigate the angular instability using different resources, for example, synchronous condensers, wind turbines, and STATCOMs~\cite{Nolan2021}. Reduced fast dynamic reactive power support can lead to a voltage instability condition cascading into frequency instability, the so-called voltage-dip-induced frequency dip phenomenon~\cite{Creighton2013}. Consequently, \textit{fast post-fault active power recovery} is also a new AS designed to incentivize faster and sustained active power recovery of wind power plants after a fault on the system~\cite{EirGrid2020}. Black start resources are procured through bilateral contracts, or a tender, and compensated by availability payments and other additional costs in EirGrid~\cite{EirGridblack2020}. In SONI, black start capability is mandatory and remunerated by a regulated payment. 

The generic payment structure for individual services in EirGrid/SONI comprises the product of the available volume, multiplied by a regulated fixed tariff and also by a scalar. The latter includes various multipliers to reward system-friendly providers and penalize less beneficial participants, depending on, certain capabilities. For example, resources eligible to provide \textit{fast frequency response} that are capable of responding very quickly, such as BESS, within 0.15 seconds, will be paid (much) more. Participants receive higher payments if they provide services under scarcity conditions (high VRE levels), locations, and can sustain their response across consecutive reserve categories. In the case of non-delivery, payments are reduced~\cite{EirGrid2017}. In EirGrid/SONI, the costs of balancing services are recovered from consumers through a tariff.

\subsubsection{Germany}

The energy transition in Germany is supported by high subsidies for solar and wind sources, which allow individuals and private cooperatives to be self-producers, decentralizing generation~\cite{Buchan2012}. Additionally, the country phased out its nuclear power plants and designed a competitive tender process to compensate coal-fired power plant owners that deactivate their units~\cite{IEA2021}. Despite the increasing shares of VRE and the phase out of SG, Germany does not experience significant stability issues. \textit{Instantaneous reserve}, that is, an inertia emulation service, was discussed by Dena (the Germany Energy Agency) to overcome possible problems arising from low inertia levels under the ``\textit{Ancillary Services Study 2030}" ~\cite{DENA2014,DENA2018}. However, investments in transmission expansion and its favored location in the center of Europe, which provides strong AC interconnections, have lessened some system needs observed in less well-interconnected regions. A sensitive system need observed in Germany is implementing congestion management to avoid bottlenecks in all voltage levels. The country is moving from a cost-based redispatch to a market-based approach, aiming to adequately remunerate flexibility from the demand side~\cite{Hirth2019}. 

Currently, German TSOs maintain organized markets to procure frequency containment reserves (\textit{primary control reserve}), automatic and manual frequency restoration reserves (\textit{secondary control reserve} and \textit{minute reserve}, respectively). Replacement reserve is not procured in Germany. Voltage control is mandatory and only paid if agreed upon through a bilateral contract~\cite{Energiforsk2020}. Eligible black start resources receive a fixed annual payment aiming to recover costs. A tariff applied to transmission consumers reimburses operating costs.

\subsubsection{Nordic Power System}

The Nordic synchronous system (Eastern Denmark, Sweden, Finland, and Norway) is marked by high shares of hydropower, accounting for 54.5\% of the generation, while other renewables represent 15.5\%. Asynchronous links interconnect the Nordic power system with the Continental and Baltic synchronous area\footnote{The Vyborg HVDC link, which connects Finland and Russia, ended its operation after the beginning of the Ukraine war on February 2022 ~\cite{ENTSOEhvdc2022}.}. Two frequency containment reserve (FCR) products are available in the Nordic synchronous system. FCR-N is constantly maintained within the normal frequency band, while FCR-D is dimensioned to withstand disturbances when the steady-state frequency deviation exceeds 0.5\,Hz~\cite{Statnett2017,Fingrid2022}. Critical inertia levels often arise in summer at night when consumption is low and wind production is high. \textit{Fast frequency reserve} is procured to complement the FCR-D product in case of a low-inertia event to reduce the maximum frequency deviation~\cite{ENTSOE2019}. IBRs and loads capable of fast response, within one second, are traded in single clearing price capacity auctions. TSOs acquire automatic frequency restoration reserves in a regional balancing capacity market, forecasting imbalances in the bidding zones and available transmission to dimension the amount of reserve~\cite{NordicaFRR2022}. A regional balancing capacity market for manual frequency restoration reserves is under design. Instead of manually activating the reserve capacity, each TSO will determine the demand for reserve in their bidding zone, according to forecasted imbalances, thereby enabling a central optimization algorithm for offer selection~\cite{Nordic2019,Khodadadi2020}. Notice that TSOs in the Nordic power system currently do not acquire replacement reserves. Voltage control is a mandatory AS, and bilateral agreements define the remuneration of suppliers. If sufficient offers are submitted, a competitive tender procures black start resources in Denmark; otherwise, bilateral contracts are established~\cite{Elia2018}. As in Sweden, the operating costs can be shared between consumers and BRPs.

\section{United States versus European Ancillary Services Market Design}

The different market design choices of the US and Europe impact the definition of standard ASs in each region. Figure~\ref{fig:USA_vs_EU} compares the US (red bars) and European (blue bars) frequency-related products regarding their time frames. 

\begin{figure}[!h]
    \centering
    \includegraphics[width=1.0\columnwidth]{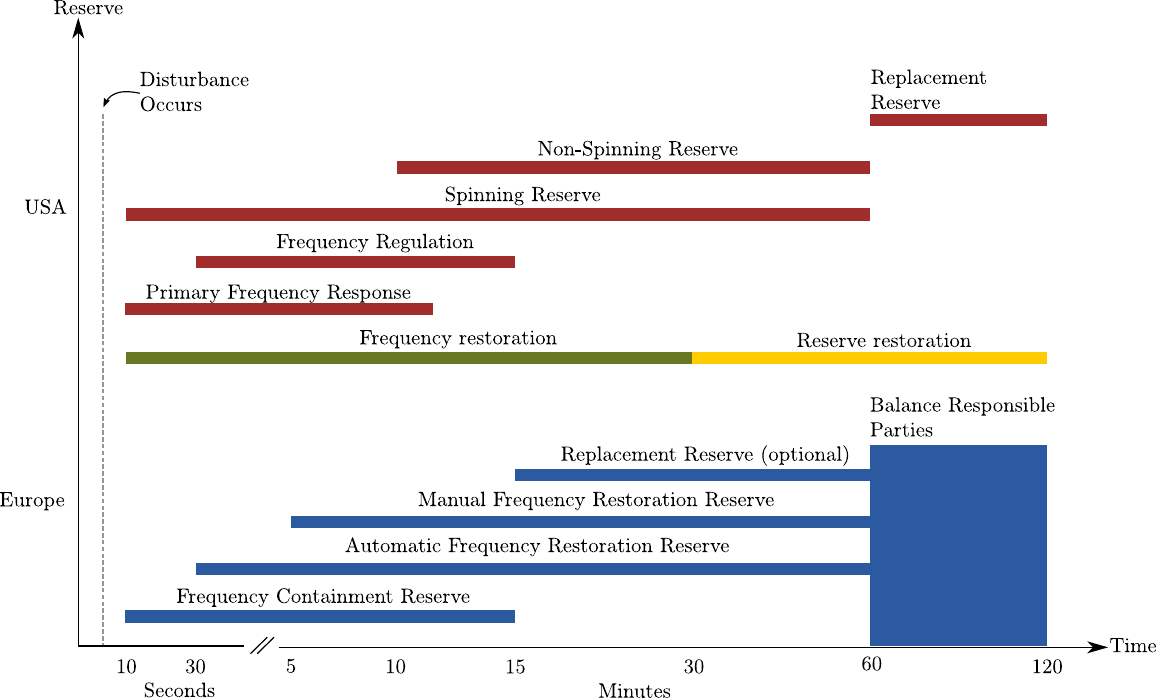}
       \caption{Standard frequency-related ASs in the US and EU wholesale electricity markets. Conventionally, frequency should be restored within 30 minutes (green bar), and subsequently, additional capacity restores the reserve levels (yellow bar).}
    \label{fig:USA_vs_EU}
    \vspace{-4mm}
\end{figure}

The requirements for primary frequency control are satisfied by primary frequency response and frequency containment reserve in the US and Europe, respectively. Under normal operating conditions, secondary frequency control (SFC) is performed by frequency regulation in US, and automatic frequency restoration reserve (aFRR) in Europe. Under contingency conditions, US defines spinning and non-spinning reserves as the standard products to perform SFC, while the European framework considers the deployment of aFRR (SFC) followed by manual frequency restoration reserve, which is related to tertiary frequency control (TFC). Both designs consider replacement reserves as an additional reserve associated with TFC to replenish reserve levels. 

Under high VRE shares, the demand for reserves tends to increase to mitigate forecast errors~\cite{Seel2018}. However, the German experience shows the opposite. Adequate market design and other factors have so far been sufficient to compensate for the increased variability and uncertainty of wind and solar generation, as highlighted in~\cite{Hirth2015}. Higher temporal granularity in the intraday market (15-minute interval) allows participants to continuously update their positions, reducing the need for balancing reserves in real time. The European balancing platforms enable exchange of the procured reserves among TSOs, optimizing the management of portfolios. Additionally, reduced security margins, infrequent outages of generators, and improved forecasting tools are also factors that have contributed to diminishing the need for balancing reserves. In ERCOT, market design changes also have reduced the procurement of frequency regulation capability despite the increasing share of wind generation. When ERCOT moved from a zonal to a nodal market, the portfolio-based dispatch was replaced by a unit-specific dispatch, which ensures more detailed control of generation. By shortening the dispatch interval, from 15 to 5 minutes, the small imbalance uncertainties between dispatch intervals were reduced, diminishing the need for frequency regulation reserves~\cite{Andrade2018}. Also, wind power plants are much faster responding than conventional generation, and thus, much tighter control can be achieved.

\section{Modern ASs Market}

Dominant VRE participation does not imply changes in physics or economics concepts, but such considerations must be respected to ensure an efficient (secure and at least-cost) power system operation. Nevertheless, new system needs are pushing AS needs into the center of electricity market design discussions. Rather than being labeled as auxiliary, services are instead essential toward a 100\% non-fossil future transition. The main technical and market design gaps are now highlighted, pointing out potential solutions or research needs to orient academic researchers and industry practitioners.

\subsection{Technical Challenges}

New system needs imposed by the displacement of synchronous resources and high VRE levels reveal frequency and voltage control shortfalls that must be mitigated. Grid-forming inverters are a promising technology, but not yet considered mature for bulk power systems. Additionally, in order to unlock the hidden potential of DERs, better prediction of future installed capacity and dynamic load modeling are needed.

\subsubsection{Power System Stability Concerns}

Two major topics are selected here: namely, frequency and voltage control issues. The main challenge for frequency stability is to contain the higher RoCoF and frequency deviation resulting from an inverter-dominant power system. For the voltage stability case, DERs should contribute to solving problems which can help avoid the need for additional grid infrastructure.   

\paragraph{Frequency Control Issues}

The proliferation of IBRs and modern loads interfaced by power electronic inverters within AC power systems might allow greater tolerance regarding the frequency variation range in the future. Nonetheless, current practice is to keep the system frequency within a narrow band around the nominal value. To tackle the reduction of system inertia, SOs could adopt modern inertia monitoring methods using data from PMU measurements, rather than dispatch-based estimation, to capture the time-varying demand-side inertia contribution~\cite{Liu2021,Farantatos2021}. Under high shares of variable generation, quantifying regional inertia is particularly important to guarantee sufficient ability for a potential island grid formed after a transmission line tripping to support frequency control~\cite{Tan2022}. Thus, more accurate estimation is helpful in determining inertia-related thresholds, enhancing coordination with inertial and fast response reserves. 

To transport higher volumes of VRE generation over long distances, the transfer capability of HVDC links tends to rise, increasing the size of the largest contingency, and worsening the higher RoCoF and lower maximum frequency deviation problems, if system inertia levels are reduced. Therefore, a greater volume of RoCoF and frequency-deviation-based reserves should be procured by SOs for frequency containment. Since reserve provision tends to shift from conventional synchronous toward inverter-based resources, and currently, virtual inertia capability is not widespread, it follows that under very high instantaneous VRE shares, limiting HVDC imports and curtailing VRE generation could help to ensure acceptable frequency deviations, but at an operational cost~\cite{OSullivan2014}.

\paragraph{Voltage Control Issues}

A reduction in online synchronous generation means that power systems will increasingly rely on capacitor banks, shunt reactors, FACTS, and IBRs to provide reactive power and maintain secure voltage stability margins~\cite{NG2021}. Since the voltage should be controlled locally, the current location of some devices could result in insufficient voltage support in some locations, requiring optimal placement strategies. Additionally, IBRs have lower fault current contributions compared to SGs, which compromises the ability of protective systems to sense and clear faults promptly, making the fault propagate through the network and potentially ending in a cascading outage~\cite{Gordon2022}. Most grid codes require that wind and solar plants remain connected and contribute with reactive power after a severe voltage drop. Ride-through capability requirements (voltage and frequency) can be further extended to DERs.

A less centralized power system also encourages enhanced participation of DERs to actively regulate voltage. In the US, utilities and ISOs are gradually incorporating the requirements proposed in the IEEE 1547-2018 standard, updated to consider high DER shares and modern inverter capabilities (active control and ride-through functionality)~\cite{IEEE2018}. For example, the gradual movement of clouds alters PV generation, resulting in higher voltage fluctuations, which can be mitigated by injecting or absorbing reactive power from modern inverters~\cite{Sun2019}. Also, during light load conditions and peak solar production, reverse power flow from the low-voltage feeder through the distribution substation can lead to voltage rise in the former~\cite{Tonkoski2012}. To accommodate the installation of new devices, network reinforcement is necessary. Alternatively, investments can be deferred using modern inverters capable of absorbing reactive power. 

\subsubsection{Grid-Forming Inverters Deployment at Scale}

Grid-forming (GFM) inverter solutions, driven by non-synchronous resources, could potentially assist power system stability and further provide low-carbon AS. The necessary requirements to extract GFM inverter benefits in bulk power systems still need to be defined in grid codes, but the technology is not widely available. To date, GFM inverter capabilities have been demonstrated mainly in microgrids and isolated systems, and grid code initiatives are limited, such as in National Grid ESO}~\cite {Lin2020}. On the other hand, without clear capability specifications and market incentives, manufacturers could be discouraged from developing the technology. This circular problem is being addressed by testing GFM inverters at the transmission level through pilot projects, enabling identification of the main barriers to GFM adoption at scale in bulk power systems~\cite{Phurailatpam2023,Matevosyan2021}.

Inverters, both GFL and GFM technologies, are physically limited by the availability (wind and solar headroom) or the size (battery capacity) of an energy buffer to respond to fast active power variations after disturbances. Thus, a key issue is to determine the reserve level that GFM-based resources should maintain to preserve system stability. GFM-based solutions should avoid trying to fully replace SG capabilities, such as high fault current, which requires greater energy buffer capacity and increases overall costs. Instead, if SOs could quantify the benefits arising from GFM inverter connection to system stability, providers could be rewarded, which would accelerate technology development, allowing higher shares of non-synchronous resources~\cite{AEMO2021}.

In weak grids, GFM inverters could improve local voltage stability and support a minimum system strength to allow the connection of additional grid-following (GFL) inverters, which are cheaper and, in the future, could potentially be converted to GFM capability~\cite{NERC2021GFM}. Also, GFM inverters can provide VIR and support frequency control. Distinct from GFL technology, GFM inverters are capable of black start. Nevertheless, how to coordinate GFM-based resources between different sites to create power islands that can restore the whole system is, to date, an open research question~\cite{NGnblack2019}. Another sensitive point is the interaction between different inverter technologies and SGs, which potentially introduces harmonics, new oscillation modes, and resonances, resulting in system instabilities that need further investigation~\cite{Denholm2021}. 

\subsubsection{Improving Distributed Energy Resources Visibility}

In order to access the benefits of integrating DERs in distribution systems, such as AS provision, these resources should be sufficiently visible for DSOs and SOs in planning studies and real-time operation. During the planning phase, DSOs should accurately estimate future DER capacity additions in the grid to avoid system security problems and increased costs. Improved DER capacity forecasts require collecting data from individual consumer characteristics, such as electricity consumption from electricity bills, suitability of rooftops from satellite data, etc. Using data-driven models (\textit{bottom-up} modeling), DSOs can weight consumer characteristics against potential DER insertion, providing better visibility of future DER capacity. However, detailed consumer data may well not be readily available, requiring significant investments in data acquisition. Also, economic uncertainties, such as future capital costs and DER regulatory policies, and modeling uncertainties, introduced to simplify consumer dynamics, are inherent shortfalls that should be addressed carefully, potentially creating multiple solution scenarios~\cite{Gagnon2018,Horowitz2019}.

New system needs, such as enhanced local voltage control and congestion management, arise due to increasing shares of DERs. SOs should consider a dynamic equivalent of emerging active distribution networks to analyze the effects of many hidden modern loads on system stability, and improve operation planning~\cite{NERC2017}. Currently, neither SOs nor specialized software companies have a sufficiently versatile tool to perform detailed transient stability simulations of modern loads~\cite{IEEE2022}. The key issue is to provide a sufficiently accurate and flexible load model, capable of being generalized under different operating conditions and locations, with reasonable computational time to allow integration with SO tools. A potential solution, based on research findings, is developing a \textit{gray-box} model, which considers the load composition and system dynamics measurement information~\cite{Rabuzin2022}.

\subsection{Market Design Challenges}

Potential market design issues arising from high VRE shares can be aggregated as two fundamental goals of electricity markets: (\textit{i}) efficient price signals, particularly real-time price signals, and (\textit{ii}) competition. The most relevant issue regarding (\textit{i}) is scarcity pricing. Supply shortages should be reflected in real-time prices, and propagated through long-term decisions to ensure reliability, since high shares of VRE are shifting the revenue streams from energy toward flexibility (ASs) and capacity. On the other hand, TSO-DSO coordination is the most important reform to be implemented regarding (\textit{ii}). To explore flexibility from distribution-based resources and promote competition between large and small players, improved coordination between the TSO and DSO(s) is a crucial point. 

\subsubsection{Improving Real-Time Pricing Signals}

In US, relevant issues to enhance price formation, such as multi-interval pricing and non-convexities, emerge from the need to align the optimal dispatch instructed by the ISO with the profit-maximizing objectives of flexible resources. European balancing market redesign could focus on co-optimizing reserves and improving locational and temporal pricing signals in real time. Although explicit scarcity prices are primarily addressed in US, the topic is also sensitive for price formation in Europe.   

\paragraph{Multi-Interval Dispatch and Pricing}

The need for increasing operation flexibility requires suitable consideration of the intertemporal constraints, such as ramping rates, start-up cost allocation, and ESS dynamics, including battery state-of-charge and hydrogen volume tank level, to avoid distorted price signals. Ramping constraints, for instance, tend to bind more often in real time under high VRE levels, which has led to the creation of a specific AS to reward the opportunity costs to dispatch out-of-merit units. However, ramping products, such as those proposed in CAISO and MISO, may not obtain the least-cost operation solution~\cite{Schiro2017}. An alternative approach solves the real-time economic dispatch and pricing looking ahead to future time intervals through a rolling horizon. Using the projected power system conditions, ISOs can pre-position resources to manage forecasted binding ramping constraints~\cite{Hogan2021}. NYISO and CAISO have implemented \textit{multi-interval pricing} in their RTM, while in ERCOT, the approach remains in the proposal stage. 

Nevertheless, the advisory prices emerging from multi-interval pricing when a rolling horizon is considered cannot support the optimal economic dispatch because there is no financial or physical commitment after the first interval. By acting as rational profit-maximizing agents, flexible resources are encouraged to self-dispatch, deviating from ISO instructions. Therefore, side payments (or uplift payments) should be provided, even in a convex market, to preserve a consistent market outcome. In particular, dispatch supporting prices can be provided if a fixed horizon and perfect foresight are considered~\cite{Hua2019}. The lack of commitment to future prices could be fixed by a financially binding look ahead, which would be updated and modified at each successive interval. The financial commitment at each set of look-ahead adjustments preserves the dispatch incentives and avoids the need for uplift payments.

\paragraph{Revenue Insufficiency due to Non-Convexities}

For the US ISOs, uplift payments are also needed in the presence of non-convexities arising from fixed costs, such as start-up and no-load costs, and minimum generating power constraints from the SCUC in order to make resources whole, since marginal pricing fails to recover non-convex costs. Since uplift payments suppress price signals, market transparency is undermined, which may lead to inefficient operating and investment decisions if a significant volume of payments are provided~\cite{ONeill2005}. Variable generation imposes more frequent cycling (start-up and shut-down) of conventional generators to accommodate stochastic net load fluctuations, which can increase total uplift payments~\cite{Wolak2021}. Introducing new ASs, such as synchronous inertial response, includes additional non-convexities in the unit commitment problem, since units may be committed out-of-merit exclusively to provide this service~\cite{Viola2023}. Also, a non-convex market for voltage support based on the solution of the AC power flow could efficiently remunerate providers, compared to cost-based methodologies, encouraging them to support voltage stability. \textit{Convex-hull pricing} is theoretically the preferred solution to tackle non-convexities and reduce the lost opportunity cost, a certain type of uplift payment. However, solving the Lagrangian dual problem is currently computationally impractical for real power systems~\cite{Andrianesis2022}. Alternatively, a computationally efficient primal implementation that closely approximates convex-hull prices is proposed in~\cite{Hua2017}. Also, another approximation, commonly called \textit{approximate extended} LMP (aELMP), which relaxes the integrality of binary variables, is currently used in MISO and PJM.

\paragraph{Transition to Co-Optimization of Energy and Reserves}

The European DAM design considers separate markets and entities to procure electricity (power exchanges on the energy market) and reserve capacity (TSOs on the balancing capacity market) to sequentially clear energy and reserves. Sequential optimization leads to a sub-optimal dispatch of energy and reserve capacity, which prevents the best use of available resources. To avoid a potential lack of reserves in real-time operation, forward reservation capacity typically occurs well ahead of real-time electricity activation in Europe. The current approach implies that participants should infer their future imbalances, which is becoming increasingly difficult under high shares of VRE, to embed an estimation of their opportunity cost in their offer~\cite{Baldick2017}. Inefficient allocation of reserves may result in a price reversal condition, whereby providers of low-quality reserves receive higher compensation than better-quality providers, resulting in disincentives for flexible resources~\cite{Eronen2022}. Joint optimization of energy and reserves explicitly reflects the opportunity cost of holding back reserves instead of generating electricity, resulting in efficient price signals and avoiding the need for redispatch actions. However, several institutional obstacles make the transition to simultaneous optimization of energy and reserves a complex task in Europe. Some issues include integrating power exchanges and TSOs platforms, and the resulting impacts on EUPHEMIA performance. Also, the ongoing harmonization of balancing services across Europe is crucial~\cite{Smeers2021}.

\paragraph{Increasing Locational and Temporal Price Granularity}

Accurate real-time price signals are essential to reflect the uncertainties in real-time operation introduced by high VRE shares. Fundamentally, balancing energy markets manage energy deviations between day-ahead and real-time operation, considering day-ahead prices as a reference. Thus, price formation in Europe strongly relies on DAM, which harms the consistency with RTM. Price formation centered on real-time prices allows better use of the balancing energy exchanged in ongoing European platforms. Additionally, two possible refinements are increasing locational and temporal price granularity.

Moving towards nodal pricing may optimize the use of flexible resources due to their improved visibility, thus avoiding redispatch actions or curtailment of VRE, and so, reducing total operating costs. Moreover, manipulative bidding in market-based redispatch, so-called \textit{inc-dec gaming}, which aggravates congestion, can be eliminated under the nodal pricing approach~\cite{Eicke2022}. Nodal prices could arise from an RTM with co-optimization of energy and reserves~\cite{Antonopoulos2020}. Besides balancing product harmonization, electricity market governance should be discussed to avoid conflict interests in transmission line use~\cite{Ashour2021}. Harmonization of the imbalance settlement period from 60 to 15 minutes is an ongoing directive in Europe, but it remains far from the 5-minute US ISOs standard. Moving from continuous trading to more frequent auctions in the intraday market could encourage the participation of smaller providers that cannot invest in sophisticated trading mechanisms to improve the speed of trades~\cite{Ehrenmann2019}. Furthermore, shortening gate closure times, allows participants to more accurately update their positions near real-time operation, potentially reducing the need for reserves.

\paragraph{Explicit Scarcity Prices}

Considering the increasing levels of near-zero marginal cost generation, short-term price signals (electricity and reserves) should adequately induce efficient long-term investment decisions. If available capacity is tight, prices should rise to reflect supply scarcity. However, administrative price caps and weak demand participation require the introduction of an explicit scarcity component in short-term pricing~\cite{Joskow2019}. An Operating reserve demand curve (ORDC) is a mechanism to determine real-time reserve capacity prices, and the associated scarcity adder in real-time electricity prices, to optimally allocate capacity between electricity provision and reserve for system security~\cite{Hogan2019}. The ORDC approach was first implemented in ERCOT and is now also adopted by PJM and MISO. In US, all other ISOs are proposing reforms to include improved scarcity pricing~\cite{FERC2021}. The authors in~\cite{Papavasiliou2021} investigated the inclusion of the ORDC approach under the European market design due to the lack of an RTM for reserve capacity. The explicit scarcity prices provided by an ORDC, incentivize operational flexibility from controllable loads, DERs, and fast-ramping resources to be available during shortage conditions, fostering long-term investments in new flexible capacity~\cite{Hogan2022}. 

Important improvements in the ORDC modeling should be considered. The traditional static model based on historical information could evolve into a dynamic model based on updated information available from forecasting tools~\cite{Lavin2020}. By coupling probabilistic forecast methods and stochastic programming models, multiple dispatch scenarios could indicate a more conservative reserve procurement~\cite{Tuohyesig2019}. Also, the ORDC approach could be expanded to address multiple reserves and locations. All the aforementioned improvements are hard to be implemented and have computational barriers~\cite{Hogan2019}. 

\subsubsection{Removing Barriers for Competition}

Market arrangements should facilitate a technology-agnostic provision and the entry of low-carbon AS suppliers in the market, enabling competition among resources with diverse cost structures and availability, irrespective of their voltage level. In the following, inefficient market arrangements, such as symmetrical offers and renewable subsidies, are first shown to prevent some participants from competing in the electricity markets. Afterward, better TSO-DSO coordination is discussed as an enabler for improving DER visibility and competition in electricity markets.  

\paragraph{Inefficient Market Arrangements}

In some US and European markets, offers for services, such as frequency regulation, require a symmetrical reserve capability in both upward and downward directions to accept providers. Upward capability involves a greater opportunity cost than downward capability, in most circumstances, except for low load and high renewable conditions. However, variable generation typically operates at its maximum power point without sufficient headroom for increased supply. Also, renewable subsidies, widely adopted to incentivize power system decarbonization, can further exacerbate the problem of higher opportunity costs of upward reserve supply from VRE, discouraging the AS provision~\cite{vanderWelle2021}. Thus, a single product for upward and downward frequency regulation inhibits wind and solar power plants from making offers to provide this service\footnote{Similarly, conventional generators operating at their minimum power cannot offer upward regulation if VRE levels increase suddenly.}. Separate upward and downward products would better reflect power system conditions, enabling more efficient use of resources~\cite{Kahrl2021}. A transition from energy- (feed-in tariffs) toward capacity-based subsidies (through technology-specific auctions), which competitively support installed capacity, rather than current energy production, can improve operational decisions~\cite{Newbery2018}. Also, the introduction of an explicit market-based price for carbon emissions can provide investment signals for low-carbon sources, reducing the need for subsidies. 

Another common barrier for VRE and ESS is the minimum capacity required by SOs for service provision, which precludes individual DERs from accessing the market. Reducing the minimum capacity itself may be insufficient to incentivize individual small and dispersed resources to compete against large units. Nevertheless, if DER aggregation is allowed, competition is enhanced. In Europe, the imbalance pricing method is an additional barrier. TSOs have widely applied the dual pricing scheme for the financial settlement of imbalances. In this case, if a balancing responsible party (BRP) faces a negative individual imbalance (shortage), it must pay the balancing service costs plus a penalty. 
Penalizing negative imbalances incentivizes BRPs to over-contract reserves in the DAM to financially hedge against the risk of being short in real-time operation. Large market participants are favored since they can strategically manage their portfolio to settle imbalances, discouraging the participation of small players~\cite{Meeus2020,Chaves2014}. A transition to a single pricing scheme, in which negative and positive imbalances are settled at the same price without penalties, would incentivize flexible resources to balance the power system. Germany, the Netherlands, and Belgium have already implemented a single imbalance pricing scheme, while France and the Nordic power system plan to do so. Addressing the foregoing market design inefficiencies is an important step to making AS markets a viable source of revenue stream for low-carbon providers. 

\paragraph{Improved TSO-DSO Coordination}

Decentralization of the power system has transferred reserve capacity from the transmission to the distribution system. If reserve capacity from the transmission level to preserve system security is insufficient, SOs should procure reserves from the distribution level. To avoid misaligned actions, closer real-time coordination between DSO and TSO/ISO is essential to leverage a bottom-up AS provision. Individual or aggregated resources in the distribution system should alter their operating plans to offer reserve capacity for balancing the transmission system. Also, active management of distribution constraints by the DSO could impact transmission system balance~\cite{EURELECTRIC2013}. Thus, fundamental issues are determining the roles and responsibilities of TSO and DSO, and who has precedence in using DER capability. 

The set of activities in which system operators should be involved depends on the TSO-DSO coordination model. Joint optimization of local (DSO) and common (TSO-DSO) AS markets, for instance, involves a close interaction between system operators, aiming to determine the least-cost solution to match transmission and distribution system needs~\cite{Gerard2016}. In 
fully centralized models, the TSO has priority to use DER capability. In contrast, purely local AS markets are managed by a DSO, which has precedence to reserve capacity from dispersed resources. Alternatively, a decentralized common AS markets defines priority to use DER capability through the combined solution of the local and common markets. The local market is cleared first with distribution grid constraints, but without the commitment of units. Afterwards, the common market is cleared, considering the previous solution and the transmission constraints~\cite{Gerard2018}. Including local AS markets reduces computational and information exchange complexity compared to a fully centralized model, while reaching near-optimal allocation~\cite{Papavasiliou2018}.  

\section{Conclusion}

As power systems transition to higher shares of VRE, new system needs are directing a review of existing AS suites toward a 100\% non-fossil future. New frequency-related AS have been recently defined to mitigate reduced levels of system inertia, ensure fast-acting reserves, and promote flexible ramping capability. Emerging services related to voltage control focus on maintaining system stability under contingency conditions. Although increasing participation of inverter-based resources is one of the roots of stability problems, they are evolving and are also an integral part of the solution. TSO-DSO coordination is fundamental to extracting DER flexibility, allowing competition with large players.
The inclusion of explicit scarcity prices ensures efficient real-time prices and adequate long-term investment signals by addressing supply shortage needs. Particularly in US, better allocation and pricing of ramp capability could be achieved by adopting multi-interval dispatch and pricing. Also, improving price formation to handle non-convexities could reduce the financial losses of more frequent start-up and shut-down operations. In Europe, joint optimization of energy and reserves would optimally allocate and price available resources. Although a transition to nodal pricing and further refinement of temporal granularity could be challenging, the potential benefits of improving resource visibility, and thus, achieving more efficient price signals, should be considered.   

\singlespacing
\bibliographystyle{elsarticle-num}
\bibliography{ref}

\end{document}